# Discovery of a hybrid topological quantum state in an elemental solid

**Authors:** Md Shafayat Hossain[1]*†, Frank Schindler[2,3]*†, Rajibul Islam[4,5]*, Zahir Muhammad[6]*, Yu-Xiao Jiang[1]*, Zi-Jia Cheng[1]*, Qi Zhang[1], Tao Hou[7], Hongyu Chen[7], Maksim Litskevich[1], Brian Casas[8], Jia-Xin Yin[1], Tyler A. Cochran[1], Mohammad Yahyavi[7], Xian P. Yang[1], Luis Balicas[8], Guoqing Chang[7], Weisheng Zhao[6], Titus Neupert[9], M. Zahid Hasan[1,10,11]†

**Affiliations:**

[1]Laboratory for Topological Quantum Matter and Advanced Spectroscopy (B7), Department of Physics, Princeton University, Princeton, New Jersey, USA.

[2]Blackett Laboratory, Imperial College London, London SW7 2AZ, United Kingdom.

[3]Princeton Center for Theoretical Science, Princeton University, Princeton, NJ, 08544, USA.

[4]International Research Centre MagTop, Institute of Physics, Polish Academy of Sciences, Aleja Lotników 32/46, PL-02668 Warsaw, Poland.

[5]Department of Physics, University of Alabama at Birmingham, Birmingham, Alabama 35294, USA.

[6]Hefei Innovation Research Institute, School of Microelectronics, Beihang University, Hefei 230013, P.R. China.

[7]Division of Physics and Applied Physics, School of Physical and Mathematical Sciences, Nanyang Technological University, Singapore.

[8]National High Magnetic Field Laboratory, Tallahassee, Florida 32310, USA.

[9]Department of Physics, University of Zurich, Winterthurerstrasse, Zurich, Switzerland.

[10]Princeton Institute for Science and Technology of Materials, Princeton University, Princeton, NJ, USA.

[11]Lawrence Berkeley National Laboratory, Berkeley, California 94720, USA.

†Corresponding authors, E-mail: mdsh@princeton.edu; f.schindler@imperial.ac.uk; mzhasan@princeton.edu.

*These authors contributed equally to this work.

**Abstract**
**Topology and interactions are foundational concepts in the modern understanding of quantum matter. Their nexus yields three significant research directions: competition between distinct interactions, as in the multiple intertwined phases, interplay between interactions and topology that drives the phenomena in twisted layered materials and topological magnets, and the coalescence of multiple topological orders to generate distinct novel phases. The first two examples have grown into major areas of research, while the last example remains mostly untouched, mainly because of the lack of a material platform for experimental studies. Here, using tunneling microscopy, photoemission spectroscopy, and theoretical analysis, we unveil a "hybrid" and yet novel topological phase of matter in the simple elemental solid arsenic. Through a unique bulk-surface-edge correspondence, we uncover that arsenic features a conjoined strong and higher-order topology, stabilizing a hybrid topological phase. While momentum-space spectroscopy measurements show signs of topological surface states, real-space microscopy measurements unravel a unique geometry of topology-induced step edge conduction channels revealed on various forms of natural nanostructures on the surface. Using theoretical models, we show that the existence of gapless step edge states in arsenic relies on the simultaneous presence of both a nontrivial strong $Z_2$ invariant and a nontrivial higher-order**

**topological invariant, providing experimental evidence for hybrid topology and its realization in a single crystal. Our discovery highlights pathways to explore the interplay of different kinds of band topology and harness the associated topological conduction channels in future engineered quantum or nano-devices.**

## Main text

Despite its dubious reputation as the "king of poisons," arsenic (As) has long been used in semiconductor technology, tonics, and even for cancer treatment [1]. Here we find that the most stable form of As, namely α-As [2], hosts a novel quantum phase of matter which, at low energies, can be understood via the mathematical framework of topology [3,4]. Prototypical examples of this class of materials are topological insulators, which act as an insulator in their interior but feature time-reversal symmetry-protected conducting states on the surface or edge [3-7]. In particular, strong (or first-order) topological insulators in three-dimensional crystals are gapped in the bulk but stabilize gapless surface states, while higher-order topological insulators are also gapped on their surfaces but host one-dimensional gapless hinge modes [8-10]. The controllability of such topologically protected conduction channels has garnered immense interest in energy-saving technology and quantum information science. By combining first- and higher-order topology in a single material, this work provides a route to such precise control of a single crystal quantum system. Leveraging a combination of experiments and theory, we find that α-As carries simultaneous topological modes on the surface and along the step edge(s). Notably, we show that the step edge modes appear only for selected, crystalline-symmetry-allowed orientations, which – unlike surface or hinge states – are not expected for either first- or higher-order topological insulators, but only for a hybrid material where both kinds of band topology are present.

The discussion of the electronic structure of α-As is facilitated by the presence of inversion symmetry, which allows us to designate the bands at high-symmetry momenta with inversion eigenvalues. The number of "+" and "-" eigenvalues below the Fermi level reveals the so-called band inversions in momentum space and links to symmetry-indicator topological invariants [11,12]. α-As features a series of band inversions around the Fermi level that combine to produce its hybrid topology [13-15]: (i) A double band inversion at the Γ point of the rhombohedral (primitive) Brillouin zone, which, on its own, would result in a higher-order topological insulator with gapped surfaces but helical hinge states [12]. (ii) A single band inversion at the three $C_3$-related *L* points of the rhombohedral Brillouin zone (See Extended Fig. 1**a**), which is expected to lead to a first-order topological insulator[11] with gapless Dirac cone surface states. Existing angle-resolved photoemission spectroscopy work on α-As set out to explore the first-order topological insulator state [13], reporting topological Rashba-type, spin-split surface states. However, a real-space investigation into how the higher-order topology is intertwined with the first-order topological insulator phase remains lacking. Owing to high spatial resolution, electronic detection, and magnetic field tunability, scanning tunneling microscopy provides a means to visualize the topological boundary modes directly in real space [16-30]. By revealing gapless step edge-localized states, our scanning tunneling microscopy measurements expose the unique, hybrid first- and higher-order topology of α-As.

We start with the surface characterization of α-As. The crystalline lattice has a rhombohedral primitive cell with the space group of $R\bar{3}m$ (No. 166). In the following, we work with the hexagonal (conventional) cell for convenience. As illustrated in Fig. 1**a**, the crystal can be regarded as a stack of As bilayers along the *c*-axis. The As atoms within the bilayer form a buckled honeycomb lattice, and cleaving leaves a triangular lattice in the *ab* plane (Fig. 1**b**). Using scanning tunneling microscopy on a freshly cleaved sample, we visualize a large surface area shown in Fig. 1**c**, featuring atomically resolved triangular lattice (Fig. 1**c** top inset) with sharp Bragg peaks in the Fourier transform image (Fig. 1**c** bottom inset). Figure 1**d** summarizes our energy-resolved tunneling spectroscopy measurements on a clean region, taken over a large distance. The averaged spectrum reveals a soft gap, i.e., a suppression in the spectral weight near the Fermi energy. In Fig. 1**e,** we show the temperature-dependent resistance data measured in a bulk α-As sample while passing the current along the *ab* plane. Here, the resistance decreases as the sample is cooled down, indicative of a metallic state on the surface. Indeed, the presence of such a surface state is captured by our angle-resolved photoemission spectroscopy measurements. Figure 1**f**, where we plot the second derivative of an energy-momentum cut along the $\bar{\Gamma} - \bar{M}$ direction of the surface Brillouin zone (Extended Fig. 1**a**), provides such data. We observe a pair of sharp, parabolic surface bands that is consistent with both prior photoemission work [13] and our first-principles calculations (as shown in Fig. 1**g**), indicating the presence of a topological Rashba-like surface state (see also, Extended Figs. 1 and 2) along $\bar{\Gamma} - \bar{M}$. Note that the bulk spectrum, when projected on the surface Brillouin zone, is gapped along the $\bar{\Gamma} - \bar{M}$ line (Extended Fig. 1). Hence, the surface state within the bulk gap marks the topological bulk boundary correspondence, indicating a first-order topological insulator phase in α-As. Additional photoemission spectroscopy and first-principles calculation results pertaining to the topological surface state in α-As are summarized in Extended Figs. 1 and 2.

Notably, the surface state also manifests in our real-space imaging. Our approach invokes the quasiparticle interference in the *ab*-plane. Extended Fig. 3, which captures the results of such experiments, shows a prominent scattering branch whose origin and shape match the parabolic surface band visualized in our photoemission spectroscopy (Fig. 1**f**) and first-principles calculations (Fig. 1**g**); see Extended Table I for a quantitative comparison between the three sets of data. Thus, quasiparticle interference spectroscopy provides a real-space detection of the presence of a surface state in α-As.

To gain insight into the surface state more quantitatively, we investigate the Landau quantization in α-As via tunneling spectroscopy measurements under perpendicular magnetic fields and construct a Landau fan diagram. We note that mapping out the Landau fan through tunneling experiments is a non-trivial task, and there are only a few successful examples in the world of quantum materials, including ultra-clean materials such as graphene [31], bismuth [32], $Bi_2Se_3$ [33], PbSnSe [34], and $TbMn_6Sn_6$ [29]. Here, in α-As, as we slowly increase the magnetic field, a plethora of states emerge, manifested via tunneling differential conductance (d$I$/d$V$) peaks that are widely distributed in energy (Fig. 1**h** top panel); this is a clear signature of Landau quantization. As a function of energy and magnetic field, the tunneling spectra form a Landau fan diagram shown in Fig. 1**h** (see Extended Fig. 4 for details). A close examination of Fig. 1**h** data reveals two sets of electron-type Landau fans. These two sets of Landau fans can be fitted to the Rashba-split up- and down-spin Landau levels (marked with color-coded dashed lines in Fig. 1**h**) stemming from the Rashba-split surface state in α-As. Here, the terms 'spin up' and 'spin down' refer to two sets of electronic structures with opposite spin directions, without specifying any particular direction. Thus, we obtain quantitative band structure information for the Rashba-like surface state such as an effective mass of 0.11 $m_e$ and Rashba spin-orbit coupling strength of 0.6 eV· Å (see section VII.B in Methods for details). It is also worth pointing out that the two sets of Landau fans converge to the same bottom energy that closely matches the bottom of the surface band visualized in the photoemission (Fig. 1**f**), first-principles calculations (Fig. 1**g**), and quasiparticle interference data (Extended Fig. 3**c**). Overall, our observation of a Rashba-split Landau fan provides comprehensive evidence for the presence of Rashba surface states in α-As.

Having characterized the topological surface state in α-As, we turn to investigate the electronic structure of its step edges. We performed d$I$/d$V$ spectroscopic maps at a monolayer step edge along the *a*-axis direction identified by a topographic image (Fig. 2**a**). We find that the step edge exhibits pronounced step edge states within the soft gap (spectroscopic images in Fig. 2**a**). Figure 2**b** captures the line profile obtained from the d$I$/d$V$ map, demonstrating an exponential decay (with a characteristic length of $r_0 \simeq 2.1$ nm) of the step edge state on the crystal side. The step edge state decays more sharply on the vacuum side. The corresponding topographic line cut is presented in Fig. 2**c**. Further spectroscopic measurements taken perpendicular to the step edge (Fig. 2**d**) reveal a peak in the step edge spectra (orange curves) near the Fermi energy (at $\simeq -5$ mV) where the spectra away from the step edge (violet curves) exhibit a soft gap.

A topological edge state is protected by the time-reversal symmetry, which can be broken by applying an external magnetic field. When we apply a magnetic field perpendicular to the *ab* plane, we find that the differential conductance measured at the step edge is suppressed substantially. The field-induced suppression of the step edge state can be clearly seen in the d$I$/d$V$ map taken at $B = 2$ T, shown in Fig. 2**e**. Upon further examination of the field-dependent tunneling spectra (Fig. 2**f** for $B = 1$ T, 2 T, and 4 T), we find that an energy gap gradually develops at the step edge state. The response of the step edge state to an out-of-plane magnetic field is consistent with the established phenomenon in time-reversal symmetric materials, whereby the field induces the mixing of helical edge states and the opening of a Zeeman gap [35-37]. Although we observe well-developed Landau levels in α-As (Fig. 1**h**), it is a semimetal with a significantly higher carrier density compared to typical semiconductor quantum wells [35]. Consequently, the magnetic fields employed in our experiments do not suffice to trigger strong quantum Hall edge states, for example, the integer quantum Hall state at Landau level filling factor 1, at the step edge. The magnetic field essentially opens a gap at the step edge state Kramers degeneracy point. The energy gap position in Fig. 2**f** suggests that the Kramers degeneracy point is located around $-5$ mV, slightly below the Fermi energy. From our field-dependent data, we estimate the gap opening rate of $\sim 2$ meV/T, suggesting a large *c*-axis Landé *g*-factor of $\sim 35$ for the edge. Note that, the large *g*-factor, which may arise from the orbital contributions [38], has also been reported in other topological materials [30, 38, 39]. Overall, the observed gapless edge state resides in an energy window of the low surface density of states and develops a gap in response to a time-reversal symmetry breaking perturbation, pointing to a protection by time-reversal symmetry as is known from the topological edge or hinge states [6].

To investigate the quasiparticle interference of the step edge state and compare it to that of the surface state, we performed energy- and spatially-resolved d$I$/d$V$ spectroscopies perpendicular (top panel of Fig. 2**g**) and along (top panel of Fig. 2**h**) a monolayer step edge. In both cases, we observe pronounced ripple-like quantum interference patterns. By taking a one-dimensional Fourier transformation of the spectroscopic data with respect to the spatial location, we obtain quasiparticle scattering patterns perpendicular (bottom panel of Fig. 2**g**) and along (bottom panel of Fig. 2**h**) the step edge. The quasiparticle interference perpendicular to the step edge captures the parabolic dispersion of the surface state centered at $q = 0$ which is akin to the data shown in Extended Fig. 3**c**. In contrast, the quasiparticle spectrum along the step edge exhibits a single, sharp scattering branch covering a distinct energy-momentum space, which is different from the scattering branch of the surface state shown in the bottom panel of Fig. 2**g**. Remarkably, this branch comes to an abrupt halt at finite momentum transfer with a distinct finite slope – a feature that cannot arise solely from a one-dimensional band dispersion. Instead, it aligns well with a band structure in which a step edge band hybridizes with surface states at a finite momentum. A comparison between the edge quasiparticle interference (bottom panel of Fig. 2**h**) and the edge d$I$/d$V$ spectra (Fig. 2**f**) reveals that the energy location of the d$I$/d$V$ peak in the latter seems to coincide with the point of the dispersive branch vanishing in the former. Additional insights derived from a straightforward two-dimensional, lattice-based model are discussed in Sec. SIII of the Supplementary Information. Notably, this contrasts with bismuth[19], where (*i*) the step edge scattering does not vanish at finite momentum and (*ii*) the edge d$I$/d$V$ peaks were associated with singularities of the density of states at the top and bottom of the edge bands.

Having examined the energy-momentum relationship of the monolayer step edge state, we proceed to explore step edges with different geometric configurations. After extensively searching the sample surface, we found two bilayer step edge geometries along the *a*-axis identified by the topographic images in Fig. 3**a**, **b**. The bilayer step edge in Fig. 3**a** exhibits a pronounced step edge state manifested via a d$I$/d$V$ peak at $V \simeq -5$ mV (d$I$/d$V$ maps in Fig. 3**a** and spectroscopy in Fig.

3**c**), resembling its monolayer counterpart. The dispersion of the bilayer step edge state is shown in Extended Fig. 5. We observe a single prominent scattering branch that is qualitatively similar to that of the monolayer step edge state displayed in Fig. 2**h**. In sharp contrast, the d$I$/d$V$ spectrum taken on the step edge in Fig. 3**b** (d$I$/d$V$ maps in Fig. 3**b** and spectroscopy in Fig. 3**c**) features a suppressed spectral weight near the Fermi energy. We can understand this discrepancy by noting that the two step edge geometries shown in Figs. 3**a** and 3**b** are not related to each other by any crystalline symmetry. In Fig. 3**a**, the sample extends to the left of the step edge, while in Fig. 3**b** it extends to the right. Correspondingly, even though both step edges are aligned along the same crystal axis, they do not exhibit the same electronic structure (the case for monolayer step edges is demonstrated in Extended Fig. 6). This situation is reminiscent of the hinge states in higher-order topological insulators, which only appear on alternating edges along the same axis, as for instance seen in bismuth[40]. Interestingly, however, a pure (non-hybrid) higher-order topological insulator would not exhibit gapless monolayer or bilayer step edge states – the step edge height would need to be thermodynamically large to guarantee gapless hinge states (shown in Extended Fig. 7). At the same time, a pure (non-hybrid) strong topological insulator is expected to generically only feature surface states on the plateaus bordering the step edge, rather than well-localized gapless states along the step edges themselves. Below and in the Supplementary Information, we explain how the step edge states measured in α-As instead result from hybridization between the gapless first-order surface states and the gapped higher-order hinge state precursors. They, therefore, represent an immediate experimental indicator for hybrid topology. Notably, this contrasts with bismuth, where our *ab-initio* calculations (Supplementary Information Sec SIV) reveal a gapless edge mode in monolayer bismuth, aligning with previous experiments[19]. Conversely, for monolayer α-As, our *ab-initio* calculations indicate the presence of a pair of gapped edge states (Supplementary Information Sec SIV). The emergence of gapless edge modes in α-As step edges occurs only upon hybridization with the surface states existing in the bulk step edge geometry.

The concept of the hybrid topology is further orchestrated in a triangular pit (visualized in Fig. 3**d**) composed of a bilayer step edge along the *a*-axis direction and two monolayer step edges along the *b*-axis and rotated by 120º with respect to it. Owing to the same geometric configuration as that of the step edge in Fig. 3a, the bilayer step edge features a pronounced step edge state, as highlighted in d$I$/d$V$ maps (Fig. 3**d**) and differential spectrum at the bilayer step edge (Fig. 3**e**). It is worthwhile emphasizing that the bilayer step edge in Fig. 3**d** shares similar electronic properties with the monolayer step edge in Fig. 2. First, the step edge spectrum features a peak at $V \simeq -5$ mV, akin to the monolayer step edge and the bilayer step edge in Fig. 3**a**. Furthermore, in response to the time-reversal symmetry breaking, the step edge state develops an energy gap as a function of the magnetic field that is evident from the tunneling spectra under magnetic fields (Fig. 3**f**). Correspondingly, the differential conductance at the step edge becomes substantially suppressed; this is captured in Fig. 3**g**, which show the d$I$/d$V$ map in the presence of a 2 T magnetic field. Intriguingly, however, the monolayer step edges in Fig. 3**d**, which are oriented along the *b*-axis and rotated by 120º with respect to it, do not show an enhancement in the d$I$/d$V$ along the step edge. In fact, as is evident in Fig. 3**e,** instead of a peak in the d$I$/d$V$ spectra, they exhibit a suppressed spectral weight near the Fermi energy, pointing to the absence of a step edge state along these two step edges. Analogous to the absence of step edge states along certain bilayer (monolayer) orientations which are clear in Fig. 3**b (**Extended Fig. 6), this finding suggests that one should observe step edge states at the monolayer step edges along the *b*-axis (and axes rotated by 120º with respect to it) when the sample extends in the opposite direction compared to the monolayer step edges in Fig. 3**d**. Our experimental results presented in Extended Fig. 8 confirm this.

To understand the unusual geometric properties of the step edge states as discussed in the above two paragraphs, we performed systematic theoretical calculations that are detailed in the Supplementary Information. We relate the presence of step edge modes in α-As (schematically pictured in Fig. 4**a**) to its bulk topology. First-principles calculations predict that α-As is a first-order topological insulator with Dirac cone surface states [13], but also that it should host higher-order hinge states [14, 15]. We have already discussed the experimental evidence for first-order topology. In terms of the experimental signature of higher-order topology, we observe a pronounced gapless state localized at a six-atomic layer thick step edge of a specific orientation, while no such state is present in the other orientation (see Extended Figs. 9**e**, **f**, **i**). In this case, the step edge height is sufficiently thick to permit well-separated hinge states between the top and bottom hinges. Therefore, the observation of such geometry-dependent hinge states on a thick step edge can be regarded as evidence for the presence of nontrivial second-order topology. We note that, we cannot exclude the possibility of hybridization between the edge and surface states, even for higher step edges, as long as it does not lead to the gapping out of the step edge states. However, we would like to emphasize that this does not impact our conclusion, which is the presence of higher-order topology from the gapless hinge states observed on high step edges. Furthermore, we show in detail in the Supplementary Information that the presence of gapless states on thin step edges, such as the monolayer and bilayer step edges discussed above, is another unambiguous indicator of higher-order topology. This is because such states would not be present in a non-hybrid first-order topological insulator. To confirm this, we first derive the presence of gapless (helical) step edge states on general grounds, and then corroborate our analysis by an explicit tight-binding model. While the existence of gapless states in the step edge geometry only relies on α-As being a (first-order) topological insulator (as illustrated in Fig. 4**b**), their hybridization with the higher-order hinge states (detailed in Figs. 4**c, d, e**), which are finite-size gapped along monolayer and bilayer step edges, localizes them to the step edges and significantly increases their spectral weight around the Fermi level (see Fig. S1 and Sec. SI C of Supplementary Information for details). The resulting step edge electronic structure in α-As (illustrated in Fig. 4**e**) therefore hosts a helical pair of gapless states, which are protected to remain gapless by the nontrivial first-order topology in the presence of time-reversal symmetry but localized along the step edge by the concurrent nontrivial second-order topology. For the tight binding model, which has the same topological invariants as α-As, we chose idealized monolayer (Fig. 4**f**) and bilayer (Fig. 4**j**) geometries that preserves the crystalline inversion symmetry. The resulting band structure features gapless step edge states along both the monolayer and bilayer step edges, as shown in Figs. 4**g** and **k**, respectively. Consistent with the situation in α-As, the states that are closest to the Fermi level prefer one step edge orientation over another, and the preferred orientation is swapped when going from the monolayer (see Figs. 4**h, i** for orientation-resolved density-of-states) to the bilayer (see Figs. 4**l, m** for the orientation-resolved density-of-states).

Such an even-odd effect in the step-edge state preferred orientation persists in three- and four-layer step edges (shown in Extended Figs. 9**a-d**, **g**, **h**), which also display geometry-dependent step edge states. It is important to distinguish between the topologically enforced features of the step edge states and those properties that are not topological but rather arise from the microscopic details of α-As. In particular, along any given direction and for any number of layers terminating at the step edge, the existence of a single helical step edge mode for one fixed orientation (with the sample extending either to the left or right of the step edge) is topologically enforced. However, which orientation is preferred by the step edge state is not dictated by topology but rather depends on the specific properties of the material. (The same holds true for hinge modes in a higher-order topological insulator.) To decide which orientation is preferred by the step edge for a given number of layers, thereby explaining the observed even-odd effect, we need to take into account more detailed information about α-As beyond its bulk topological invariants. In fact, the orientation dependence of step edge states and the associated even-odd effect can be attributed to the asymmetry in the atomic structure between the two orientations of the step edge, which is specific to the crystal structure of α-As (see also Section XIII of Methods and Supplementary Information). In Extended Fig. 10, we further illustrate this asymmetry by showing the crystal structure and corresponding topographic images of both the monolayer and bilayer step edges. Both the monolayer and bilayer step edges display an asymmetry between the two orientations, with one orientation having a sharper step edge than the other. Interestingly, this preferred orientation alternates between the mono- and bilayer cases, in accordance with the even-odd effect observed in the step edge states, where the step edge state always prefers the smoother edge. In fact, this asymmetry simply comes about due to the buckling of the As monolayers, as explained in the Supplementary Information. (Note that solely based on experiments, we cannot isolate the effects of the layer number and the type of buckled atoms.) There, we also provide further schematic illustrations for three- and four-layer step edges to confirm that this purely geometric effect, as well as its associated mode localization, persists for thick step edges. As discussed above, a pure higher-order phase generically has gapped hinge states on a monolayer and bilayer step edge (see also Extended Fig. 7). Only in the hybrid case will the step edge state be gapless helical states that connect valence and conductance bands (see Fig. 4). Although distinguishing between these two cases based solely on the observation of step edge states may be difficult, our angle-resolved photoemission spectroscopy data presented in Fig. 1**f**, in conjunction with the merging of the surface state into the step edge state observed in the quasiparticle interference spectra in Fig. 2**g**, **h**, suggests that the hybrid interpretation is more appropriate. This observation holds crucial significance for future applications in microelectronics and spintronics utilizing α-As step edge states, as only gapless helical states enjoy protection from backscattering by time-reversal symmetry.

The combination of theoretical and experimental results presented here clearly indicates that the observed step edge states in α-As – unlike surface or hinge states – are not expected for either first or higher-order topological insulators separately, but only for hybrid materials where both kinds of band topology are present. Further theoretical modelling should aim to shed light on the soft surface gap observed in Fig. 1**d**. This soft gap – while not of topological origin – appears to be an important feature of the α-As band structure that our current theory cannot predict.

Our discovery of a hybrid topology in arsenic reveals the presence of topological step edge modes along specific geometrical configurations that are compatible with its crystalline symmetries, offering a new pathway to engineer topological step edge conduction along selective crystallographic orientations and advance quantum technologies. Furthermore, the identification of elemental solids as material platforms, such as antimony hosting a strong band inversion topology [41] or bismuth hosting a higher-order topology [40], has led to the significant development of novel materials that have immensely benefited the field of topological materials. We envision that arsenic, with its unique topology, can serve as a new platform at a similar level for developing novel topological materials and quantum devices that are not currently accessible through bismuth or antimony-based platforms.

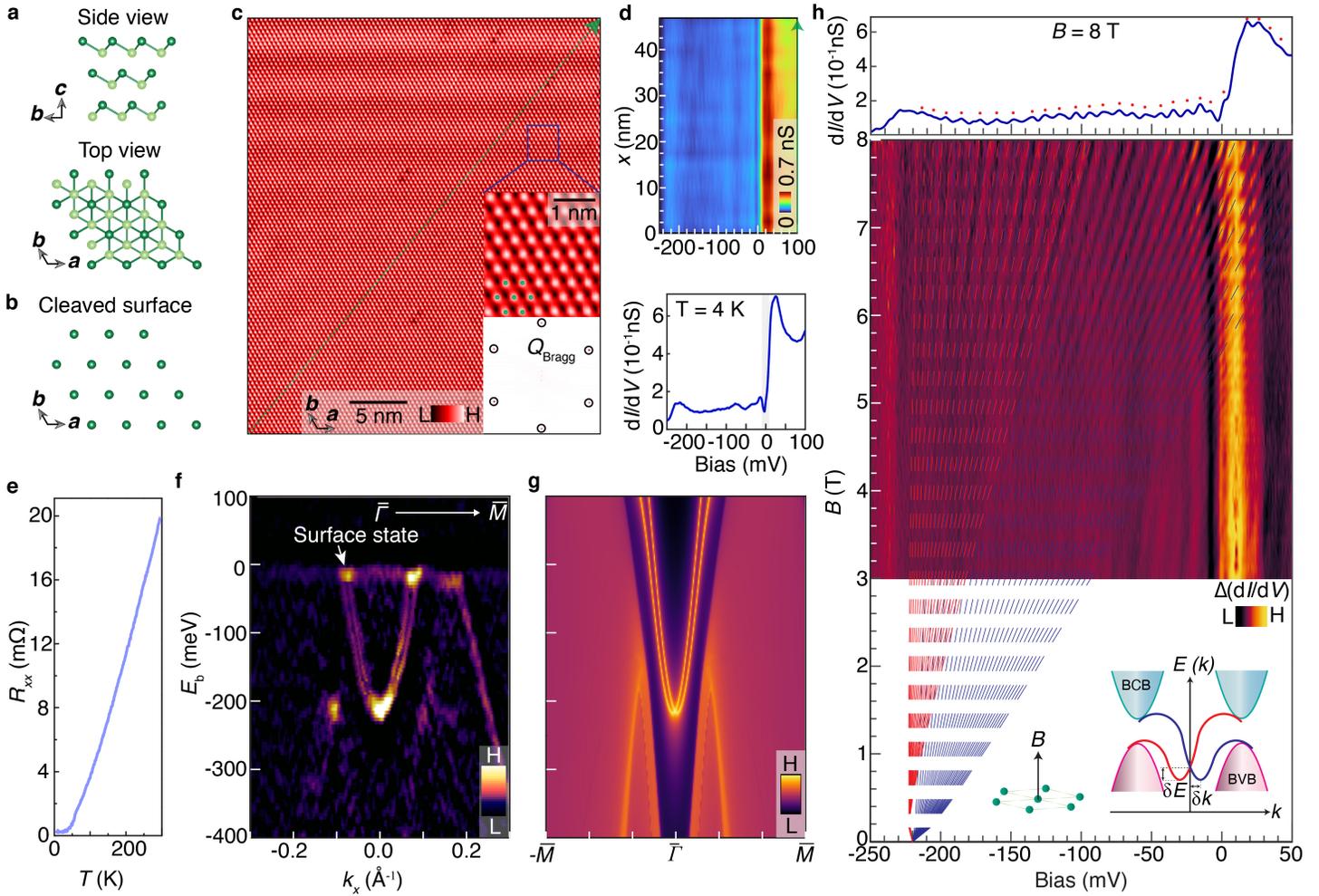

**Fig. 1. Observation of surface state in α-As and its Landau quantization. a,** Top: Side view of α-As crystal, constituting a stacking of As bilayers along the *c*-axis direction. Bottom: Top view of the crystal structure showing the As atoms within the bilayer forming a buckled honeycomb lattice. **b,** Cleaved *ab* plane comprising a triangular lattice. **c,** Scanning tunneling microscopy image of the α-As *ab* plane ($V_{gap}$ = 100 mV, $I_t$ = 3 nA). Top inset: Magnified, atomically resolved image of the triangular lattice. Bottom inset: Fourier Transform of the topographic image revealing sharp Bragg peaks corresponding to the triangular lattice. Here, the *a* and *b* axes represent two symmetry equivalent axes that are rotated by 120 degrees with respect to one another. The axis-direction illustrated here is consistently maintained throughout the manuscript to designate these axes. H and L labels in the color bar denote high and low scale, respectively. **d,** Large-scale spectroscopic linecut on the surface, taken at $T = 4$ K. The corresponding location is marked on the topographic image in panel **c** with a dotted green line; the direction of the scan is marked with an arrow. The averaged differential spectrum, shown at the bottom, reveals a soft gap (reduced spectral weight) near the Fermi energy. Tunneling junction set-up: $V_{set} = 100$ mV, $I_{set} = 0.5$ nA, $V_{mod} = 0.5$ mV. **e**, Resistance as a function of the temperature measured in a bulk α-As crystal with the contacts placed on the *ab* plane. Consistent with the presence of a metallic surface state, the resistance decreases as the sample is cooled down. **f**, Second derivative (curvature) of the energy-momentum cut along the $\bar{\Gamma} - \bar{M}$ direction of the surface Brillouin zone (see Extended Fig. 1**a**), obtained from photoemission spectroscopy on the cleaved (111) surface (*ab* plane). The spectra are acquired using 22 eV, linear-horizontally polarized light. The topological surface state is marked by the white arrow. The small Rashba splitting is visualized in the curvature plot. **g**, Calculated surface bands connecting the bulk conduction and valence bands with Rashba-like features, depicted as bright orange curves projected into the (111) surface (*ab* plane). **h**, Top: Averaged d$I$/d$V$ spectrum under $B = 8$ T applied perpendicularly to the cleaved surface, exhibiting intense modulation in differential conductance due to Landau quantization. Bottom: Landau fan diagram of α-As obtained via tunneling spectroscopy; see Extended Fig. 4 for more details. Two sets of Landau fans, corresponding to the Rashba-split surface state, are visible. The red and blue dashed lines denote a fit to the Rashba-split, up- and down-spin Landau levels, respectively. Left inset: Magnetic field applied perpendicularly to the triangular As lattice. Right inset: Schematic depiction of Rashba-split bands connecting the bulk conductance bands (BCB) and bulk valence bands (BVB). Tunneling junction set-up: $V_{set} = 50$ mV, $I_{set} = 0.5$ nA, $V_{mod} = 0.5$ mV.

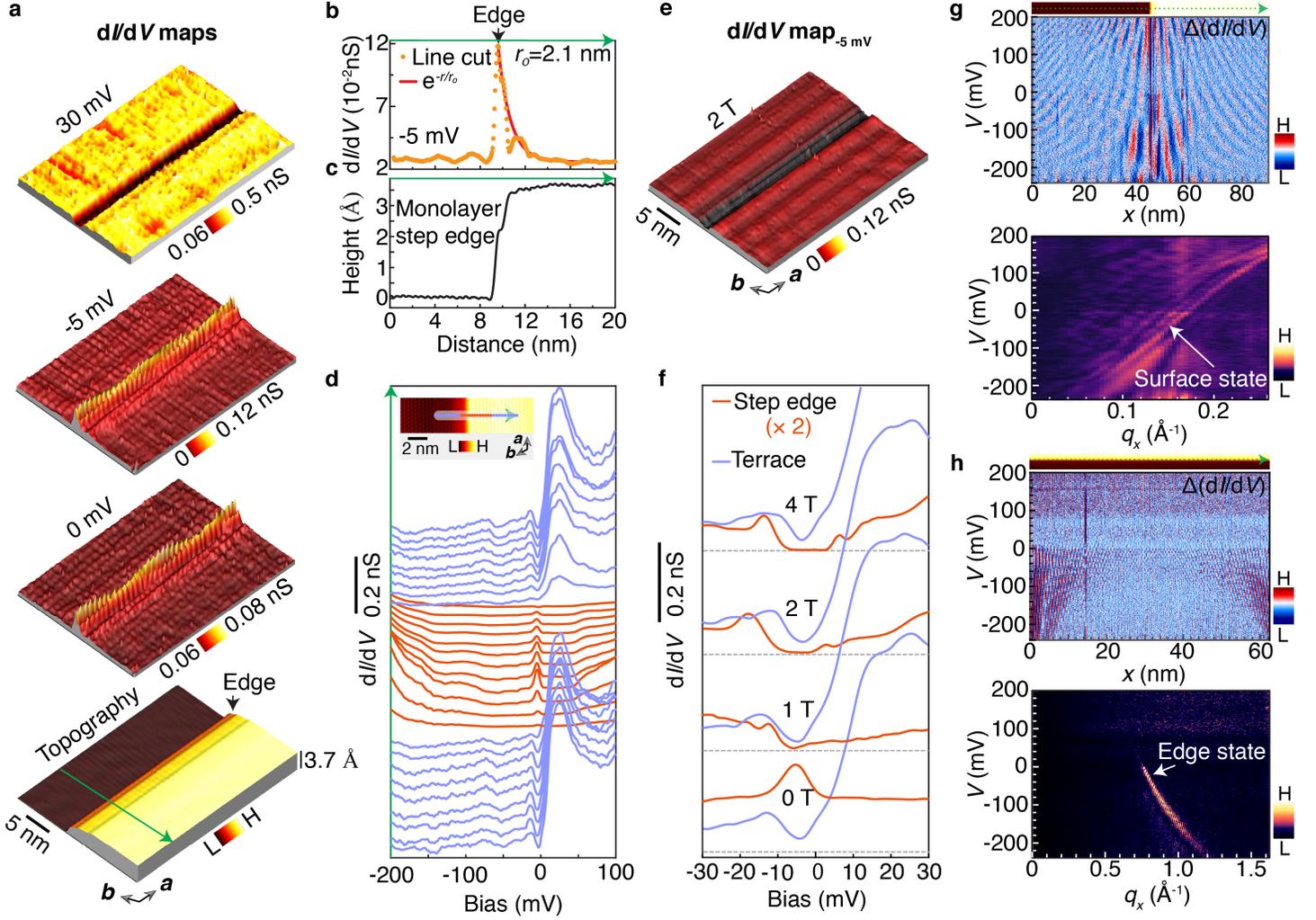

**Fig. 2. Observation of topological step edge state in a monolayer step edge. a**, d$I$/d$V$ maps at different bias voltages (corresponding topography is shown in the bottom panel) around a monolayer step edge parallel to the *a*-axis direction. d$I$/d$V$ maps taken within the soft gap near the Fermi energy ($V$ = 0 mV and -5 mV) reveal a pronounced step edge state. In contrast, at $V$ = 30 mV, the step edge state is suppressed. **b**, Intensity distribution of differential conductance taken at -5 mV around the step edge. The corresponding location is marked on the topographic image in panel **a** with a green line; the direction of the scan is marked with an arrow. The red curve shows the exponential fitting of the decay of the state away from the step edge. The fitted decay length is 2.1 nm. **c**, Corresponding height profile, taken perpendicular to the *a*-axis direction. **d**, Tunneling spectra revealing a soft gap away from the step edge and a pronounced in-gap state on the step edge. Orange and violet curves represent the differential spectra taken at different positions (marked on the topographic image shown in the inset) at the step edge and away from it, respectively. Spectra are offset for clarity. **e**, d$I$/d$V$ map at $B$ = 2 T ($V$ = -5 mV) taken in the same region as in panel **a** (topography in the bottom image of panel **a**) exhibiting a suppressed differential conductance along the edge. **f**, Differential spectra at $B$ = 0 T, 1 T, 2 T, and 4 T. Orange and violet curves denote the spectra taken at the step edge and far away from the step edge, respectively. Spectra at different magnetic fields are taken at the same locations and are offset for clarity. Dashed horizontal lines mark the zero d$I$/d$V$ for different fields. At $B$ = 1 T and 2 T, the step edge state is clearly suppressed, and an energy gap gradually develops at $V \simeq -5$ mV as a function of the magnetic field. The spectra are presented within the bias range of [-30 mV, 30 mV]. To highlight the emergence of the energy gap at the step edge as a function of the magnetic field, the step edge d$I$/d$V$ values have been multiplied by a factor of 2. This enhances the visibility of the gap in the spectra and helps to emphasize its behavior as a function of the magnetic field. The field-dependent data sheds light on the effect of time-reversal symmetry breaking on the step edge state. **g**, **h**, Line spectroscopies with high spatial and energy resolutions taken perpendicular to a monolayer step edge (panel **g**) and along the same monolayer step edge (panel **h**). This monolayer step edge, whose orientation matches that of the step edge shown in panel **a**, exhibits a clear step edge state. The top panels depict intensity plots, with green dotted lines marking the corresponding locations on the topographic images (which are shown at the top) and arrows indicating the scan directions. These plots reveal pronounced quantum interference patterns. The bottom panels display corresponding one-dimensional Fourier transforms showcasing clear dispersions. Line spectroscopy perpendicular to the step edge reveals the dispersion of the surface state, which is akin to the data obtained from a quasiparticle interference experiment performed on the surface (as shown in Extended Fig. 3**c**). In contrast, line spectroscopy along the step edge reveals the dispersion of the edge state. Tunneling junction set-up for d$I$/d$V$ maps in panels **a** and **e**: $V_{set}$ = 100 mV, $I_{set}$ = 0.5 nA, $V_{mod}$ = 1 mV. Tunneling junction set-up for the differential spectra in panels **d** and **f**: $V_{set}$ = 100 mV, $I_{set}$ = 0.5 nA, $V_{mod}$ = 0.5 mV. Tunneling junction set-up for the line spectroscopies in panels **g** and **h**: $V_{set}$ = 200 mV, $I_{set}$ = 0.5 nA, $V_{mod}$ = 0.5 mV.

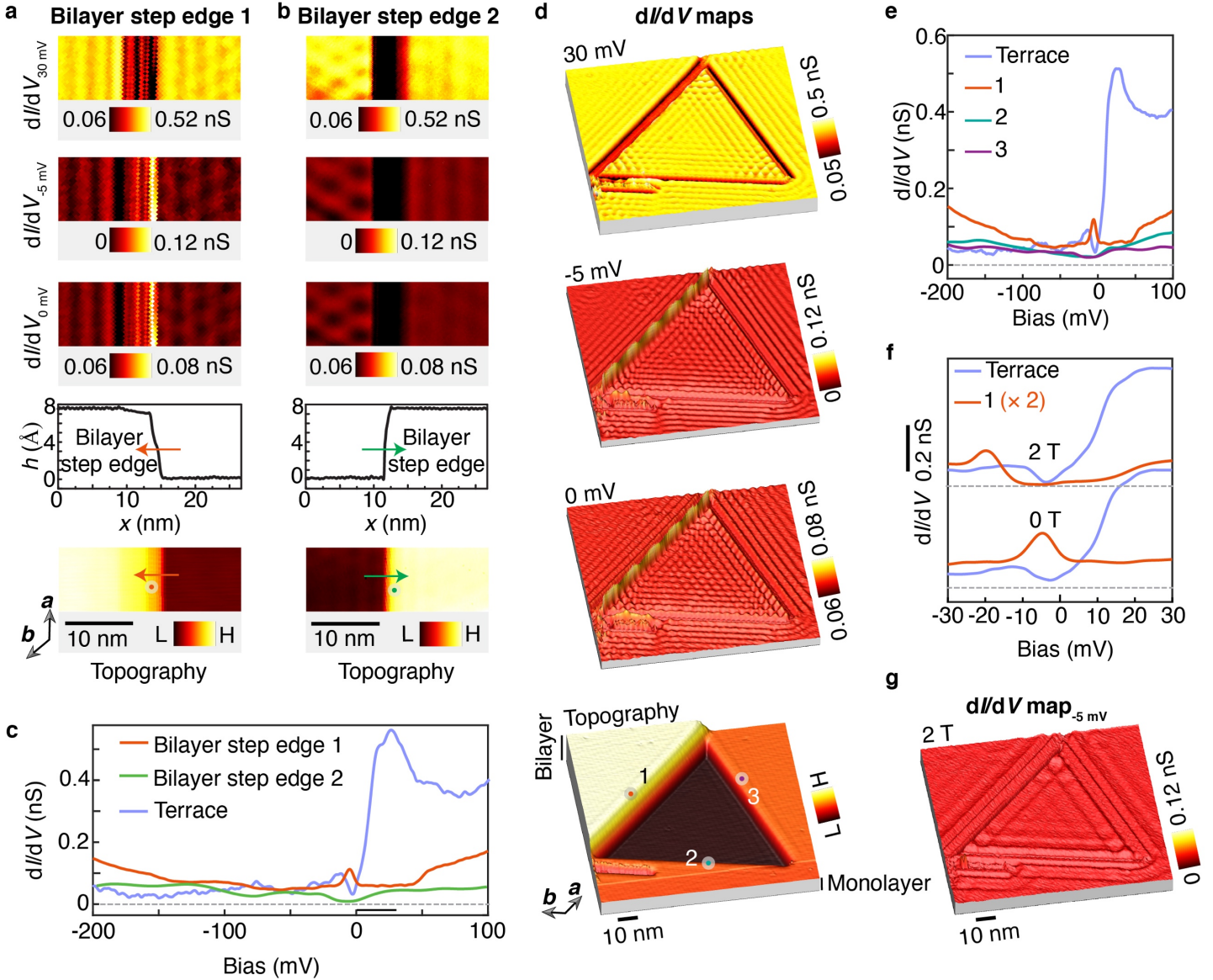

**Fig. 3. Hybrid topology: Orientation-dependence of the step edge states. a, b,** Topographic images, height profiles, and the corresponding differential conductance maps around two bilayer step edges (1 and 2 as marked). The color-coded arrows in the topographies and the height profiles indicate the directions from the bottom to the top terraces. Depending on this direction, one bilayer step edge (bilayer step edge 1) exhibits a step edge state while the other bilayer step edge (bilayer step edge 2) does not. **c,** Differential spectra, taken at the bilayer step edge 1 (orange), bilayer step edge 2 (green), and away from the step edges (violet), reveal striking differences between the two step edges. Orange and green dots in the topographic images in panels **a** and **b** denote the respective positions on the bilayer step edges 1 and 2 where the differential spectra were taken. The bilayer step edge 1 exhibits a pronounced in-gap state, whereas, on the bilayer step edge 2, the differential conductance within the soft gap is largely suppressed. **d,** Topography images and the corresponding differential conductance maps around a triangular pit composed of a bilayer step edge (whose orientation is akin to the bilayer step edge 1 in panel **a**) along the *a*-axis of the crystal and two monolayer step edges along the *b*-axis and rotated by 120º with respect to it. Like the bilayer step edge 1 in panel **a**, here also, the bilayer step edge exhibits a step edge state, whereas there are no step edge states at the monolayer step edges. **e,** Differential spectra, taken at the three step edges shown in panel **d**- bilayer step edge

(orange), two monolayer step edges (green and purple), and away from the step edges (violet). Orange, green, and purple dots in the topographic image in panel **d** denote the respective positions on the bilayer and the two monolayer step edges where the differential spectra are taken. In contrast to the bilayer step edge where a pronounced in-gap state is seen, the monolayer step edges along the *b*-axis, and rotated by 120º with respect to it, do not exhibit a step edge state. **f**, Differential spectra under $B$ = 0 T and 2 T. Orange and violet curves denote the spectra taken at the bilayer step edge in panel **d** and far away from the step edge, respectively. Spectra at the two magnetic fields are taken at the same locations and are offset for clarity. Dashed horizontal lines mark the zero d$I$/d$V$ for different fields. At $B$ = 2 T, the step edge state is substantially suppressed, and an energy gap emerges at $V \simeq -5$ mV. The spectra are plotted over a bias range of [-30 mV, 30 mV]. To illustrate the emergence of the energy gap at the step edge as a function of the magnetic field, the step edge d$I$/d$V$ values have been re-scaled up by a factor of 2. **g**, d$I$/d$V$ map at $V = -5$ mV taken on the same region as in panel **d** (topography in panel **d** bottom image) when subjected to $B$ = 2 T. The step edge state at the bilayer step edge is suppressed in response to time-reversal symmetry breaking. Tunneling junction set-up for the differential spectra: $V_{set}$ = 100 mV, $I_{set}$ = 0.5 nA, $V_{mod}$ = 0.5 mV. Tunneling junction set-up for d$I$/d$V$ maps: $V_{set}$ = 100 mV, $I_{set}$ = 0.5 nA, $V_{mod}$ = 1 mV.

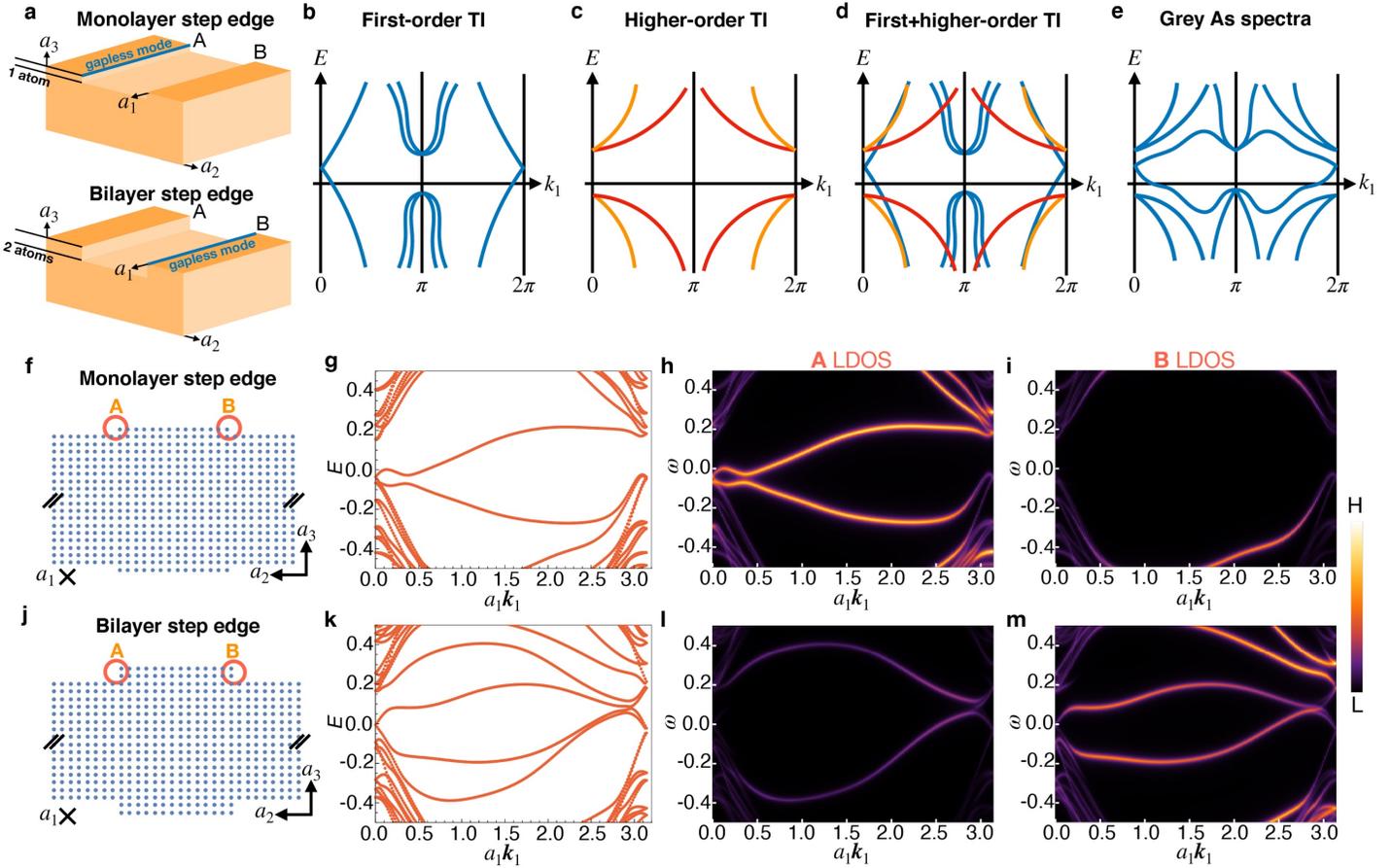

**Fig. 4. Hybrid topology of α-As manifested in the step edge state geometry. a**, Schematic depiction of the step edge states in α-As. A representative step edge geometry, which preserves one good momentum quantum number $k_1 \equiv k \cdot a_1$, is shown for both the monolayer and bilayer step edges. For comparison, we have included two step edges ("A" and "B") of opposite orientation along the same crystallographic axis. **b**, Schematic band structure of a non-hybrid first-order topological

insulator cut into the step edge geometry shown in panel a. In analogy to α-As, the valence bands are obtained from those of a trivial insulator by a single band inversion at the three $C_3$-related $L$, $L'$, and $L''$ points of the bulk Brillouin zone (see Extended Fig. 1**a**). However, unlike α-As, there is no additional double band inversion that would imply higher-order topology. As derived in Sec. I of the Supplementary Information, there is a single gapless helical mode at $k_1 = 0$, implied by the first-order topology. This state is not well-localized at the step edge but distributed across the entire surface. **c**, Schematic band structure of a non-hybrid higher-order topological insulator cut into the step edge geometry shown in panel a. In analogy to α-As, the valence bands are obtained from those of a trivial insulator by a double band inversion at the $\Gamma$- point of the bulk Brillouin zone. However, unlike α-As, there is no additional single band inversion that would imply first-order topology. There are no gapless modes, but we expect the presence of gapped higher-order hinge states that are well localized along either the A or B step edge (which step edge is chosen is a material-dependent property that is not determined by the bulk topology). In the limit of a thermodynamically high step edge (large extent in the $a_3$-direction), these gapped hinge states would become gapless. **d,** Superimposed spectra from panels **b** and **c**. **e,** Bulk electronic structure of α-As is a combination of the situations shown in panels **b** and **c**: it has a single band inversion at the three $L$, $L'$, $L''$ points, as well as a double band inversion at the $\Gamma$-point. In the step edge geometry, this leads to hybridization between the gapless surface states around $k_1 = 0$ and the gapped, step edge-localized higher-order hinge states at finite $k_1$, giving rise to gapless and step-edge localized states with an enhanced density of states close to the Fermi level. **f-m,** Step edge states in a tight-binding system modeling the band topology of α-As. **f,** Monolayer step edge geometry. We preserve periodic boundary conditions along the $a_1$-direction (the out-of-plane direction, with lattice spacing $a_1$), so that $k_1$ is a conserved crystal momentum. There are two step edges, one of type A and one of type B (see Supplementary Information for details), on each of the top and bottom surfaces. This is the minimal configuration of step edges that preserves inversion symmetry as well as periodic boundary conditions along the $a_2$-direction. **g,** Monolayer step edge dispersion with $k_1$ using the lattice shown in panel **f**. We only show the momentum range $k_1 \in [0, \pi]$ because the spectrum in the range $k_1 \in [\pi, 2\pi]$ is related by time-reversal symmetry. There is a gapless helical mode with crossing at $k_1 = 0$ (cf. panel **a**). **h,** Local density of states (LDOS) for the A step edge. In the monolayer step edge geometry, the bands closest to the Fermi level are localized along the A edge, as indicated by an increased LDOS. **i,** Local density of states for the B step edge. This LDOS vanishes for the modes closest to the Fermi energy but is large for a secondary set of bands at higher (lower) energies. Correspondingly, the lowest-energy modes are localized on the A step edge only. **j,** Bilayer step edge geometry. **k,** Bilayer step edge dispersion. **l,** Local density of states for the A step edge. The LDOS is small around the Fermi energy, indicating that there are no well-localized step edge states on the A step edge. **m,** Local density of states for the B step edge. The LDOS is large for the bands closest to the Fermi level. Correspondingly, in the bilayer step edge geometry, there are step edge modes along the B step edge. This even-odd effect between the monolayer and bilayer step edge mode orientation is consistent with the experimental findings.

## Methods:

### I.  α-As single-crystal growth

Single crystals of α-As were grown by the chemical vapor transport method. First, the As (PPM Pure Metals, 99.999995%) was filled into an evacuated quartz ampoule and was subsequently heated up to 1073 K for 48 hours. The polycrystalline As pellets were ground and filled in the quartz ampoule along with iodine shots (2 mg/cm³) as the transport agent. The quartz ampoule was then evacuated and sealed under the vacuum. The quartz ampoule was kept under a gradient of temperature zone ranging from 1073 K (growth zone) and 1173 K (source zone) for 5 days, and subsequently cooled down to room temperature at a rate of 50 K/h. Large two-dimensional high-quality α-As crystals of dimensions up to 8 mm are obtained at the low-temperature zone.

## II. Scanning tunneling microscopy

Single crystals were cleaved mechanically in situ at $T = 77$ K under ultra-high vacuum conditions ($< 5 \times 10^{-10}$ mbar), and then immediately inserted into the microscope head, already at the $^4$He base temperature (4.2 K). More than 20 single crystals have been cleaved for this study. For each cleaved crystal, we explored surface areas over 10 μm × 10 μm to search for atomic flat surfaces. Topographic images in this work were taken in the constant current mode. Tunneling conductance spectra were obtained with a commercial Ir/Pt tip (annealed in ultra-high vacuum condition and then characterized with a reference sample) using standard lock-in amplifier techniques with a lock-in frequency of 977 Hz and tunneling junction set-ups as indicated in the corresponding figure captions. The magnetic field was applied through a zero-field cooling method. To acquire the field-dependent tunneling conductance map, we first withdrew the tip away from the sample, and then slowly ramped the field to the desired value. Then we reapproached the tip to the sample, found the same atomic area, and then performed spectroscopic mapping at this magnetic field.

## III. Angle-resolved photoemission spectroscopy measurements

The angle-resolved photoemission spectroscopy measurements shown in Extended Figs. 1**d-g** and 2**d** were performed at the beamline BL21_ID (ESM) of National Synchrotron Light Source II (NSLS II). The samples were cleaved and measured at $T = 10$ K. For the Fermi surface measurements, the energy and angle resolution were below 20 meV and 0.2 degrees, respectively. For the energy-momentum cuts, the energy and momentum resolution were below 8 meV and 0.2 degrees, respectively.

The angle-resolved photoemission spectroscopy data shown in Fig. 1**f** and Extended Figs. 1**h**, **i** were performed at the beamline BL-13U at the National Synchrotron Radiation Laboratory (NSRL), Hefei, equipped with a Scienta Omicron DA30L electron analyzer. Horizontal polarization of the incident light ($h\nu = 22$ eV) was used. The energy and angular resolution were below 20 meV and 0.3 degrees, respectively. The samples were cleaved in situ at around $T = 12$ K under a vacuum better than $6 \times 10^{-11}$ mbar.

## IV. First-principles calculations

Electronic structure calculations were performed within the density functional theory framework using the Vienna Ab initio Simulation Package (VASP) [42, 43]. The General gradient approximation functional is used for treating the exchange-correlation effect [44]. A plane-wave cutoff of 500 eV was used in all calculations. The relativistic effect of spin-orbit coupling was included self consistently in the calculations. The structural parameters have been taken from experimental data. An energy tolerance of $10^{-8}$ eV was used. We have performed the calculations using 18×18×14 k-points centered at the Γ-point with 4536 *k*-points in the Brillouin zone. We extracted the real space tight-binding Hamiltonian by atom centered Wannier function with (*s*, *p*, and *d*)-As orbital projection using the Wannier90 code [45]. We constructed our model using a real space tight binding Hamiltonian to study the topological properties using the Wannier tools package [46]. The Fermi surface was calculated on the interpolated mesh of 180 × 180 × 140 using the wannier90 code and visualized with Xcrysden [47, 48].

## V. Energy-momentum diagrams using first-principles calculations and photoemission spectroscopy

Here we discuss the bulk and surface band structure of α-As in light of our first-principles calculations and photoemission spectroscopy data. The bulk and the (111) surface Brillouin zones are displayed in Extended Fig. 1**a**. Extended Fig. 1**b** captures the bulk bands, which shows a semimetal feature at the Fermi energy, i.e., both the bulk valence and bulk conduction bands cross the Fermi energy, while there is a gap between the two throughout the Brillouin zone.

Next, we turn to the electronic band structure projected on the (111) surface (*ab* plane). As shown in Extended Fig. 1**c**, the As (111) surface features a pair of parabolic bands, which split along the $\bar{\varGamma} - \bar{M}$ direction but stays degenerate at the $\bar{\varGamma}$-point (we note that these two parabolic bands form two concentric Fermi surfaces enclosing the $\bar{\varGamma}$ point (Extended Fig. 2**c**). Consistent with the prior work on α-As [13], these two parabolic bands signal the Rashba-split surface state. Upon inspecting the spin textures of the corresponding bands (Extended Fig. 1**c;** right panel), we find that the two bands are spin polarized and of opposite spins, lending further credence to the Rashba-like nature of these surface bands.

Having discussed the bulk and surface bands of α-As from the lens of first-principles calculations, we now proceed to experimentally probe these bands. In Extended Fig. 1**d-f**, we show the angle-resolved photoemission spectroscopy bulk band structure measured with synchrotron radiation. Although the energy-momentum cuts (cuts 1-3 as marked in the bulk Brillouin zone) do not completely cover the complete Brillouin zones, they still bring out the salient features of the electronic structure of α-As near its Fermi energy. Extended Fig. 1**d-f** also contains the calculated energy-momentum diagrams along the cuts 1-3 to facilitate a comparison between first-principles calculations and photoemission experiments. Note that, our synchrotron photoemission spectroscopy only captures the occupied side of the band structure. As displayed in Extended Fig. 1**d**, consistent with the first-principles results, we observe a valence band with the band maximum (near the *T* point) below the Fermi energy. On the other hand, the energy-momentum cut encompassing the $\varGamma$ point (Extended Fig. 1**e**) features no band, matching the first-principles calculations as well. Lastly, examining the energy-momentum cut along *L* (Extended Fig. 1**f**), we find that both the valence (magenta bands in the corresponding first-principles results) and conduction bands (cyan bands in the corresponding first-principles results) cross the Fermi energy. Furthermore, the spectral intensity stemming from the conduction band crossing the Fermi energy (cyan bands in the corresponding first-principles results) indicates a band minimum at the *L* point. All these band structure properties visualized in the photoemission energy-momentum cuts are in excellent agreement with the first-principles calculations.

Upon further inspection of the photoemission results presented in Extended Fig. 1**d-f**, we find that parabolic-shaped band intensities are present centering at $k = 0$ in all three energy-momentum cuts. Extended Fig. 1**g**, which depicts the photon energy dependence, further highlights the ubiquity of this band intensity throughout $k_z$ (see the spectral intensities near $k_x \sim \pm 0.05$ Å which are present in the entire range of photon energies used in our experiments). This finding is consistent with the presence of the Rashba-like surface state [13].

To carefully probe the surface state in our photoemission spectroscopy experiments, we use a lower photon energy (22 eV) and focus on a smaller energy range. Along $\bar{\varGamma} - \bar{M}$, we observe sharp, parabolic bands whose bottoms lie at E $\simeq -215 \pm$ 20 meV and close to $k = 0$ (Extended Fig. 1**h**; also shown in Fig. 1**f**). In Extended Fig. 1**i**, we present the surface state dispersion extracted from our experimental data, along with parabolic fits. Similar to the experimental bulk bands, the surface bands exhibit consistency with the first-principles results presented in Extended Fig. 1**c** (also depicted in Fig. 1**g**). We find that the separation between the Rashba-split surface bands is very small ($\Delta k \sim 0.006 \pm 0.0025$ Å$^{-1}$). This value of Rashba splitting is in excellent agreement with our first-principles results (see Extended Table I for a quantitative comparison) as well as with prior photoemission work [13], which used a high-resolution 6 eV laser as a light source and reported a similar value of $\Delta k \sim 0.0054$ Å$^{-1}$.

In summary, our comprehensive and high-quality photoemission spectroscopy measurements of the bulk states of a-As demonstrate that the first-principles calculation accurately captures the bulk states and validates our first-principles based tight-binding modeling approach. Importantly, we also confirm the surface-state nature of the Rashba state by examining the photon energy dependence of our measurements. Furthermore, our analysis of the surface-state spectrum, which is also well-reproduced by first-principles calculations, and the demonstration of the small but well-resolved Rashba splitting via synchrotron radiation based photoemission experiments not only confirm the previous laser based photoemission spectroscopy results but also provide a reference point for interpreting the Landau fan diagram and quasiparticle interference obtained via scanning tunneling microscopy measurements from the same sample.

### VI. Fermi surface mapping using first-principles calculations and photoemission spectroscopy

Here, we discuss our findings from the Fermi surface mapping via first-principles calculations (Extended Fig. 2**a-c**) and photoemission spectroscopy (Extended Fig. 2**d**). As illustrated in Extended Figs. 2**a** and 2**b**, the bulk Fermi surface in α-As contains a hole-like ($\alpha$) and an electron-like ($\beta$) pocket, both of which are three-fold symmetric. Inclusion of the Rashba-like surface states adds two concentric circular Fermi pockets centered at $k = 0$. In Extended Fig. 2**c**, we show the resulting Fermi surface projected on the α-As (111) surface. Extended Fig. 2**d** depicts the Fermi surface sheets obtained via photoemission spectroscopy on the cleaved As (111) surface. The shape of the experimentally obtained Fermi surfaces broadly match our first-principles results. Furthermore, photoemission spectroscopy measurements confirm the presence of the three Fermi pockets. The circular pocket centered around $k = 0$ (Extended Fig. 2**d**) is associated to the Fermi pockets stemming from the Rashba-like surface states.

### VII. Extended scanning tunneling microscopy data
#### VII.A. Quasiparticle interference experiment

Our systematic spectroscopic imaging on the As *ab* plane is presented in Extended Fig. 3. On a large area, we have collected differential conductance maps at different bias voltages. Thanks to the scattering of quasiparticles on the surface, the maps show real space patterns which vary as a function of bias (Extended Fig. 3**a**). By taking the Fourier transform of the d$I$/d$V$ maps, we obtain the quasiparticle interference data. At $V > -250$ mV, there is a circular-shaped signal centered at $q = 0$. The radius of the circular pattern increases as the bias is raised from -250 mV, and the pattern is clearly visible up to $V = 150$ mV. At $V = 100$ mV a six-fold-symmetric star-shaped pattern is seen inside the circular pattern which evolves at higher biases, and eventually, at $V = 200$ mV, only a six-fold symmetric line-crossing pattern is visualized. These scattering patterns exhibit a qualitative agreement with our calculated quasiparticle interference results that are obtained using the joint density of states method and presented in Extended Fig. 3**b**.

Using the quasiparticle interference data as a function of bias, we obtain a quasiparticle scattering spectrum illustrated in Extended Fig. 3**c**. The spectrum contains several quasiparticle-interference branches, the most prominent of which disperses in the form of a parabola – akin to the Rashba-like surface state visualized in the photoemission spectroscopy and first-principles calculations (Fig. 1**f**, **g**). Furthermore, the energy location of the scattering branch bottom matches the energy of the surface band bottom (Fig. 1**f**, **g**). Therefore, it is fair to say that the parabolic scattering branch in Extended Fig. 3**b** stems from the Rashba-like surface state. Note that the Rashba split band bottoms are only separated in the momentum direction (see the first-principles results in Fig. 1**g**), which is why our quasiparticle spectrum, which is not sensitive to momentum, cannot resolve it. The observed quasiparticle scattering spectrum qualitatively match the quasiparticle spectrum computed using the joint density of states method, which is presented in Extended Fig. 3d. Additionally, we provide a quantitative comparison of the quasiparticle interference spectrum with the band structure obtained from photoemission spectroscopy and first-principles calculations in Extended Table I.

#### VII.B. Mapping Landau fan diagram

Here, we discuss the measurements of the tunneling spectrum under an applied magnetic field. As elucidated in Fig. 1**d**, the zero-field spectrum is spatially homogeneous. In Extended Fig. 4**a**, we show the intensity plot of the tunneling differential conductance over a large region when the α-As sample is subjected to an 8 T magnetic field along the *c*-axis. Compared to the $B = 0$ data, the tunneling spectra at 8 T show dramatic changes with the emergence of a series of intensity peaks indicative of a series of states that are widely distributed in energy (Extended Fig. 4b). The appearance of such states signals Landau quantization. We find that the spectrum contains two nearby frequencies revealed in the Fourier transform plot (see Extended Fig. 4c), hinting at the presence of two sets of Landau fans.

To better understand the Landau quantization, we map its fan diagram in Fig. 1h by slowly ramping up the magnetic field. Mapping the Landau fan is particularly useful in extracting precise band structure information. We highlight several key features in the Landau fan diagram of α-As below.

Focusing on the Landau fans in the magnetic field range of 3 T to 8 T, we find that the Landau fans exhibit a reasonably linear dispersion as a function of the magnetic field. (We note that, at low magnetic fields, the Landau fans are not clearly visible, possibly because of thermal broadening. Therefore, for our analysis, we have used the data above 3 T.) This is consistent with the Landau fans formed within parabolic bands.

Further examination reveals that the Landau fan diagram features two sets of electron-type Landau fans. To understand this, we analytically investigate the Landau quantization stemming from the Rashba surface states of α-As, which can be described in a nearly-free-electron band dispersion [13],

$$H = \frac{1}{2m}(P_x^2 + P_y^2)\sigma_0 + \frac{g_n}{2}\mu_B B\sigma_z + \alpha(\sigma_x P_y - \sigma_y P_x), \quad (1)$$

where $\sigma_0$ is the identity matrix and $B$ denotes the homogeneous magnetic field perpendicular to the plane of motion that can be expressed as

$$\vec{B} = (0,0,B), \ \vec{A}(x,y) = \frac{B}{2}(-y,x), \quad (2)$$

In the presence of a magnetic field, the set of momentum operators is

$$\vec{P} = -i\hbar\nabla + e\vec{A} \quad (3)$$

Here we can define creation and annihilation operators

$$a^\dagger = \frac{1}{\sqrt{2e\hbar B}}(P_x + iP_y), \ a = \frac{1}{\sqrt{2e\hbar B}}(P_x - iP_y) \quad (4)$$

Then the momentum operators can be expressed in terms of the creation and annihilation operators.

$$P_x = \sqrt{\frac{\hbar eB}{2}}(a^\dagger + a), \qquad P_y = \sqrt{\frac{\hbar eB}{2}}(a - a^\dagger)i \quad (5)$$

Therefore, we can rewrite Eq. (1) as

$$H = \hbar\omega_c\left(a^\dagger a + \frac{1}{2}\right)\sigma_0 + \frac{g_n}{2}\mu_B B\sigma_z + \sqrt{\frac{2eB}{\hbar}}\alpha\begin{pmatrix} 0 & ia \\ -ia^\dagger & 0 \end{pmatrix} \quad (6)$$

The first term represents the Landau levels for two-dimensional electron gas, the second term denotes the Zeeman energy, and the third term is the Rashba spin-coupling term. Next, we define a basis that contains only the two spin states $\{|N,\uparrow\rangle, |N+1,\downarrow\rangle\}$, and the Hamiltonian in this basis is:

$$H = \hbar\omega_c(N+1)\sigma_0 - \sqrt{\frac{2eB}{\hbar}}\alpha\sqrt{N+1}\sigma_y + \left(\frac{g_n}{2}\mu_B B - \frac{1}{2}\hbar\omega_c\right)\sigma_z \quad (7)$$

Finally, we can get all the eigenvalues of the Rashba states in a perpendicular magnetic field:

$$E_{\mathcal{N}\pm} = \hbar\omega_c\left(\mathcal{N} + \frac{1}{2} \pm \frac{1}{2}\right) \mp \frac{1}{2}\sqrt{(\hbar\omega_c - g_n\mu_B B)^2 + 8\alpha^2\frac{eB}{\hbar}\left(\mathcal{N} + \frac{1}{2} \pm \frac{1}{2}\right)} \quad (8)$$

Here $\omega_c = eB/m^*$, $m^*$ is the effective mass, $\alpha$ is the strength of Rashba effect, $g_n$ is the effective g factor, $\mu_B$ is the Bohr magneton, $\mathcal{N}$ denotes the Landau levels, and $+$ $(-)$ represents spin up (down) state. It is important to note that the terms 'spin up' and 'spin down' used in this context are idiomatic and simply refer to two sets of electronic structures with opposite spin directions, without denoting specific directions. This phenomenon of Rashba-like electronic states generating two sets of Landau fans under a perpendicular magnetic field has been observed in various two-dimensional systems, including magic-angle bilayer graphene [49] and few-layer black arsenic [50]. The underlying physical mechanism of this phenomenon is that the Zeeman term induced by the magnetic field opens an energy gap at the crossing point of the Rashba electronic states (second term in Eq. 8). This leads to the Rashba electronic states splitting into two independent electronic states [51].

In α-As, existing photoemission results suggest that the Rashba states should exhibit nearly-free electron behavior [13]. Therefore, the effective $g$-factor can be set as $g_n = 1.00$, which is further corroborated in ref. [52]. (We note that the $g$-factor associated with the Rashba surface state is different from the $g$-factor that we obtained from the field-dependent edge spectra in Fig. 2.) Next, we fit effective mass ($m^*$) and Rashba effect ($\alpha$) to calculate the Landau fan. By fitting $m^*$ and $\alpha$ based on the experimentally obtained Landau fan diagram, and we obtain:

$$\text{Effective mass } m^* \simeq 0.11\ m_e$$
$$\text{Rashba strength } \alpha \simeq 0.6\ \text{eV} \cdot \text{Å}$$

Our photoemission results suggest that $\delta k \simeq 0.006\ \text{Å}^{-1}$. Therefore, a Rashba strength $\alpha = \frac{2\delta E}{\delta k} \simeq 0.6$ eV·Å leads to $\delta E \simeq 1.8$ meV; this is reasonable based on our photoemission results on the surface states. Based on the above fitted parameters, we obtain two sets of Landau fans, marked in red (up spin) and blue (down spin) in Fig. 1**h**, whose origin can be traced to $\simeq -220$ meV (We further note that the linear fits of the Landau bands do not intersect the horizontal axis at the same point because of the Rashba spin splitting). Importantly, this energy position matches the band bottom of the Rashba-like surface state obtained via a series of techniques, namely photoemission spectroscopy (Fig. 1**f**), quasiparticle interference (Extended Fig. 3**b**), and first-principles calculations (Fig. 1**g**). This observation leads us to conclude that the Landau fans in α-As arise from the Rashba-like surface state.

The Landau fan diagram also provides Fermi surface information. Taking a line cut at the Fermi energy, we obtain the field-dependent tunneling differential conductance data depicted in Extended Fig. 4**d**. The data exhibits $\frac{1}{B}$−periodic oscillations, resulting from Landau quantization of the two-dimensional electronic bands. Replotting the data as a function of $\frac{1}{B}$ and then taking the Fourier transform renders two sharp peaks, at 175 T and 187 T, as shown in the Extended Fig. 4**e**. These two peaks denote the Fermi surface area (in the units of cyclotron frequency) of the two electron-like Fermi pockets stemming from the Rashba-split surface state. Extended Fig. 4**f** schematically depicts the Rashba Fermi surface containing two concentric circular pockets. Here in α-As, the areas of the two circular Fermi pockets are 175 T (inner circular pocket) and 187 T (outer circular pocket) in the units of cyclotron frequency. We note that the two frequencies, i.e., the area of both Fermi pockets, are fairly similar. Consequently, the possible crossings between the two sets of Landau fans are not pronounced in the fan diagram shown in Fig. 1**h**.

### VIII. Quasiparticle interference along a bilayer step edge

In this section, we present the quasiparticle interference spectrum of a bilayer step edge state. Extended Fig. 5**a** depicts energy- and spatially-resolved d$I$/d$V$ spectroscopy along a bilayer step edge with the same orientation as the step edges displayed in Figs. 3**a** and 3**d**. We observe a pronounced ripple-like quantum interference pattern, which, upon one-dimensional Fourier transformation with respect to spatial location, yields the quasiparticle spectrum along the bilayer step edge. The resulting quasiparticle spectrum is shown in Extended Fig. 5**b**, where we visualize a scattering branch akin to the one shown in Fig. 2**h** for the monolayer step edge. As with the monolayer step edge state, this scattering branch is distinct from the one that stems from the surface state, as it occupies a different energy-momentum space.

### IX. Orientation dependence of the monolayer step edge in α-As:

In the main text, we have discussed the orientation dependence of the step edge states using two bilayer step edges (Fig. 3**a**, **b**), both lying along the *a*-axis of the crystal but oriented in opposite directions. These orientations differ not by the crystalline axis, but by whether the sample height increases or decreases as we cross the step edge along a fixed perpendicular direction. The data reveals that the step edge state prefers one orientation over the other, i.e., one step edge harbors a step edge state (Fig. 3**a**) while the other does not (Fig. 3**b**). In this section, we show similar data but for two monolayer step edges. Extended Fig. 6 captures such data. In Extended Fig. 6**a**, **b**, we show two monolayer step edges of opposite orientation along the *a*-axis direction, as identified by the topographic images. The monolayer step edge in Extended Fig. 6**a** does not carry a step edge state (see the d$I$/d$V$ maps in Extended Fig. 6**a**). It rather exhibits a suppressed spectral weight near the Fermi energy as seen in the spectroscopic data in Extended Fig. 6**c**. In stark contrast, a pronounced step edge state is seen in the monolayer step edge in Extended Fig. 6**b**. The step edge state is manifested via a d$I$/d$V$ peak at $V \simeq -5$ mV (see d$I$/d$V$ maps in Extended Fig. 6**b** and spectroscopy in Extended Fig. 6**c**).

Analogous to the situation where the step edge states are only seen along certain bilayer orientations, here the orientation dependence is also rooted in the fact that the two monolayer step edge geometries in Extended Figs. 6**a** and 6**b** are not related to each other by any crystalline symmetry. In Extended Fig. 6**a**, the sample extends to the left of the step edge, while in Extended Fig. 6**b** it extends to the right. Correspondingly, although the step edges are aligned along the same crystal axis, they do not exhibit the same electronic structure. Furthermore, we make the following two remarks connecting the Extended Fig. 6 data to our observations in Figs. 2 and 3. First, the monolayer step edges in Fig. 2**a** and Extended Fig. 6**b** share the same crystal axis and orientation (the sample extends to the right of the step edge). Consequently, they feature a similar electronic structure, hosting a step edge state. Second, comparing the orientation dependence of the two bilayer step edges in Fig. 3**a**, **b** with the two monolayer step edges shown in Extended Fig. 6**a**, **b**, we find that the orientation hosting gapless states is swapped when going from the bilayer to the monolayer. Specifically, for bilayer step edges, the step edge where the sample extends to the left of the step edge harbors a step edge state (Fig. 3**a**). Conversely, the monolayer step edge where the sample extends to the right of the step edge exhibits a step edge state (Extended Fig. 6**b**). As discussed in the main text (and later in Methods), such an even-odd effect appears independently of the orientation dependence of the step edge states and is related to the atomic structure of the step edges, which is determined by the crystal structure of α-As and not by its bulk topology.

### X. Extended theoretical discussion on the step edge states in α-As:

We refer to the Supplementary Information for a detailed theoretical discussion of the step edge states in a-As, as well as the derivation and Hamiltonian matrix for the tight-binding model used in our theoretical calculations.

Here we elaborate on the schematic mechanism linking the step edge states and hybrid topology that was summarized in Fig. 4**b-e**. Extended Fig. 7 captures the analog of Fig. 4**f-m** for a tight-binding model of a higher-order topological insulator that lies in the same symmetry class as α-As, but which unlike α-As does not have nontrivial first-order topology. The resulting step edge spectra, which are schematically depicted in Fig. 4**c**, are fully gapped, but do contain well-localized hinge mode precursors that contribute a significant density of states just away from the Fermi level. On the other hand, the step edge spectra of a first-order topological insulator (schematically depicted in Fig. 4**b)** generically feature a helical gapless mode, but this mode – unlike the step edge mode in the regime modelling α-As that is shown in Fig. **4f-m** – is not expected to be well localized or to contribute a significant density of states around the Fermi level.

Considering the above discussion, we conclude that the hybrid topology of α-As – a simultaneous first- and higher-order topological insulator – is indispensable for the existence of gapless step edge-localized states near the Fermi level, schematically depicted in Fig. 4**e**.

### XI. Step edge states at the monolayer step edges along the *b*-axis and rotated by 120° with respect to it

In Fig. 3**d** of the main text, we show that two monolayer step edges along the *b*-axis and those rotated by 120° with respect to it do not exhibit a step edge state. Here, in another set of monolayer step edges along the *b*-axis and rotated by 120° with respect to it but oriented exactly opposite to the monolayer step edges in Fig. 3**d**, we find pronounced step edge states. Extended Fig. 8 spectroscopically visualizes this scenario. As demonstrated in Extended Fig. 8**c**, Differential conductance maps taken near the Fermi energy (with corresponding topographic image shown in Extended Fig. 8**a**) reveal enhanced spectral weight along the step edges. This observation further highlights the orientation dependence of the step edge states in α-As.

### XII. Step edge states at three-, four-, and six-layer-thick atomic step edges

In this section, we further investigate the orientation dependence of step edge states by examining three- and four-layer thick atomic step edges, and present spectroscopic data on six-layer-thick step edges as an indication of the higher-order topological insulator phase in α-As.

We commence by presenting a spectroscopic study on two trilayer step edges (Extended Fig. 9**a**, **b**), lying along the *a*-axis of the crystal but facing opposite orientations. Our d$I$/d$V$ spectroscopies and maps reveal that the step edge where the sample extends to the right of the step edge harbors a step edge state (Extended Fig. 9**b**). In contrast, the step edge where the sample extends to the left of the step edge carries no such state (Extended Fig. 9**a**). This orientation dependence is identical to that of the monolayer step edge (presented in Extended Fig. 6). Next, we identify two oppositely oriented four-layer-thick step edges along the *a*-axis direction (Extended Fig. 9**c**, **d**). Spectroscopic measurements suggest that the four-layer step edges also exhibit orientation dependence of the step edge state; the step edge where the sample extends to the left of the step edge hosts a step edge state (Extended Fig. 9**c**) while the step edge where the sample extends to the right of the step edge features no step edge state (Extended Fig. 9**d**). This characteristic of the four-layer step edge is identical to that of the bilayer step edge but opposite to the mono- (and tri-) layer step edges. Thus, the orientation dependence of the step edge state and the even-odd effect in its preferred orientation are also evident in the case of the three- and four-layer step edges.

We have also conducted simulations of the three- and four-layer step edge geometries using our tight-binding model. Here we again take into account the asymmetry between different step edges and layer heights that is induced by the buckled structure of As monolayers (see Supplementary Information for details). We find gapless helical step edge states along both the three- and four-layer step edges, with an orientation dependence that mirrors that of one- and two-layer step edges, respectively.

Lastly, we examine two six-layer-thick step edges along the *a*-axis direction (Extended Fig. 9e, f), oriented in opposite directions. Consistent with all the examined layer thicknesses, we find that the step edge state prefers one orientation over the other. The step edge where the sample extends to the left of the step edge hosts a step edge state (Extended Fig. 9**e**), while the step edge where the sample extends to the right of the step edge has no step edge state (Extended Fig. 9**f**). Notably, the thickness of the six-layer step edge is likely sufficient to allow for well-separated and gapless hinge states between the top and bottom hinges. Because scanning tunneling microscopy is only sensitive to the top hinge, a hinge state that propagates by alternating between the top and bottom hinges would manifest as a step edge state in one orientation (where the hinge state is located on the top hinge), while the absence of a step edge state would be observed in the opposite orientation (where the hinge state is located on the bottom hinge). Therefore, our observation of such orientation-dependent step edge states on a thick step edge serves as evidence for the existence of hinge states within the higher-order topological insulator phase.

### XIII. Edge geometry of α-As

In this section, we will discuss the asymmetry in the atomic structure of α-As step edges along a certain crystallographic axis and its relation to the even-odd effect observed in the step edge states. As seen from the side view of the crystal structure depicted in Extended Fig. 10**a**, the two monolayer step edges along the *a*-direction exhibit an asymmetry, with one orientation featuring a sharper step edge. This atomic configuration can be visualized from the top view in the atomically resolved topographic images of the corresponding monolayer step edges shown in Extended Figs. 10**b** and **c**. Similarly, the bilayer step edges of two different orientations also exhibit an asymmetry, with one orientation featuring a sharper step edge (see Extended Fig. 10**d** for the schematic side view of the bilayer step edge and Extended Figs. 10**e** and **f** for the top view visualized via scanning tunneling microscopy). Notably, the orientation that features a sharper step edge alternates between the mono and bilayer cases. This switching is in accordance with the even-odd effect observed in the step edge states, where the step edge state appears to favor the smoother edge. This preference alternates between opposite orientations as the layer thickness increases from one to two layers. Such a pattern may indicate a connection between the preferred orientation of the step edge state and the crystal symmetry. It is important to note that while the bulk topology of α-As dictates the presence of step edge states in one of the two possible step edge orientations along any given crystallographic axis, it does not determine which particular orientation will host the step edge state. Here, we find that the crystal structure may play a role in this determination.

### XIV. Additional data on the step edge states

Here, we present additional data acquired on a different sample. In Extended Fig. 11, we show the topography and corresponding spectroscopic maps around a monolayer step edge (along the *a*-axis) taken at $V = -5$ mV. Consistent with the step height and the orientation of the step edge in Fig. 2, here we also observe a pronounced step edge state signaled by enhanced differential conductance at the step edge. Furthermore, complying with the field-dependent data in Figs. 2**e** and 2**f**, the d$I$/d$V$ map taken at $B = 2$ T exhibits a suppressed tunneling conductance at the edge that is consistent with the time-reversal symmetry protected nature of the step edge state.

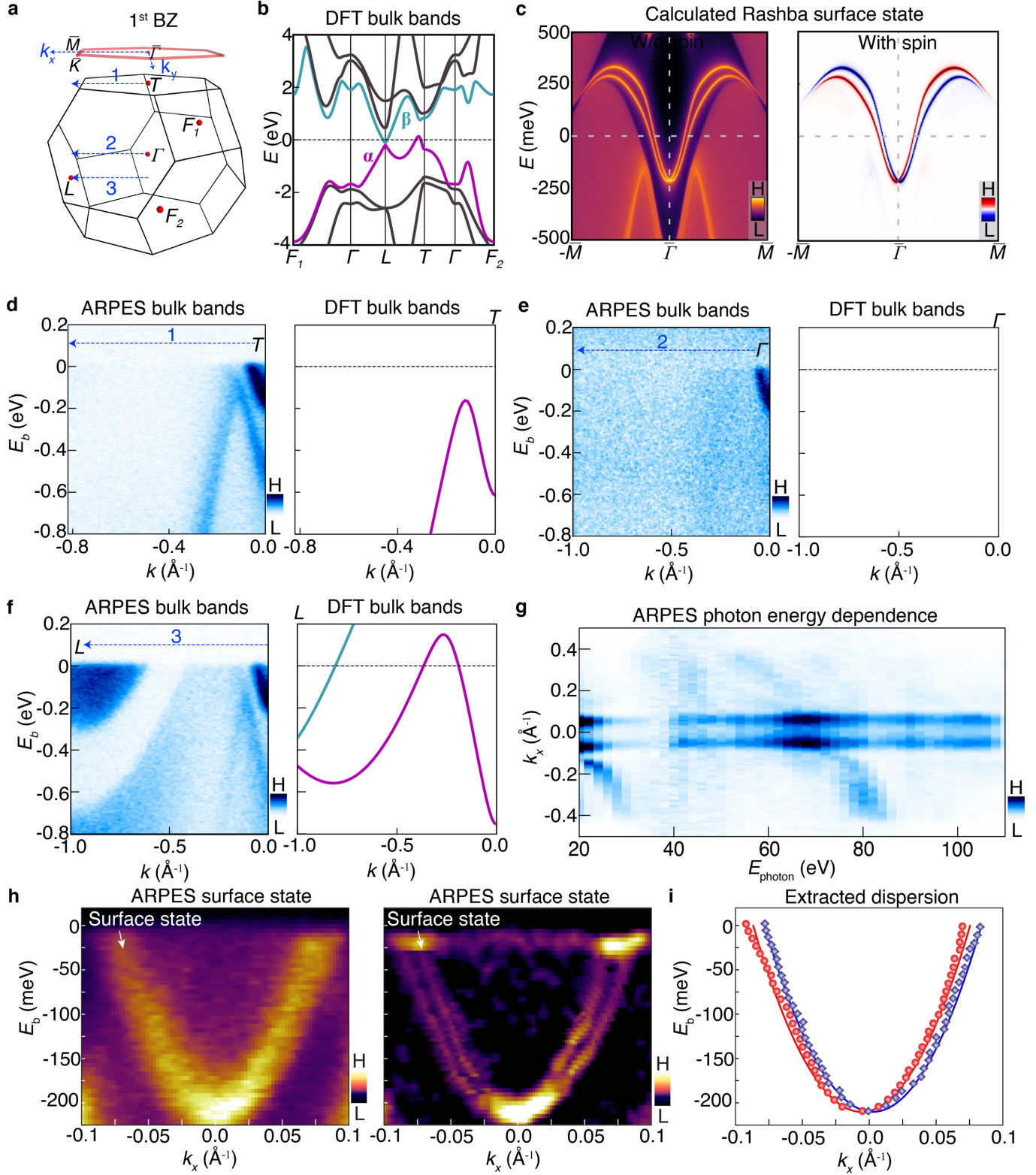

**Extended Fig. 1. First-principles calculations and angle-resolved photoemission spectroscopy determination of the electronic band structures. a,** First Brillouin zone and the projected surface along (111). Note that *L* is part of a group of three inequivalent *L*-points that are related by C$_3$ rotation symmetry around the *z*-axis. In the caption of Fig. 4 and the

Supplementary Information, these are referred to as *L*, *L'*, and *L''*. **b,** Electronic band structure of bulk α-As calculated using density functional theory over the whole bulk Brillouin zone, taking into consideration the presence of spin-orbit coupling. The bands crossing the Fermi energy are highlighted in magenta (valence band) and cyan (conduction band). **c,** Left: Calculated surface bands connecting the conduction and valence bands with Rashba-like features. These bands, depicted using bright orange curves, are projected on the (111) surface (also shown in Fig. 1**g**). Right: Calculated spin texture of the Rashba-like surface bands. **d-f,** Angle-resolved photoemission spectroscopy energy-momentum cuts encompassing high symmetry points- *T* (panel **d**), *Γ* (panel **e**), and $L_3$ (panel **f**)- of the bulk Brillouin zone. The directions are marked in panel **a** with dashed blue lines. The cuts are measured with a linear -horizontal polarized light with the photon energies of 67 eV (panel **d**), 40 eV (panel **e**), and 97 eV (panel **f**), respectively. The right panels in **d-f** show the corresponding band structures obtained via first-principles calculations. The bulk bands obtained from photoemission spectroscopy qualitatively match the first-principles results. **g**, Constant energy contour of the photon energy dependence measurement at the Fermi energy. The size of the electron pocket near *Γ* does not depend on the photon energy, which confirms its surface-state nature. **h**, Energy-momentum cut (left) and its second derivative (right) taken along $\bar{\Gamma} - \bar{M}$ direction of the surface Brillouin zone, obtained from photoemission spectroscopy on the cleaved (111) surface (*ab* plane). The spectra were acquired using 22 eV, linear-horizontally polarized light. The topological surface state is marked by the white arrow. The small Rashba splitting is visualized in the second derivative plot (also shown in Fig. 1**f**). **i**, Extracted dispersion of the Rashba-split surface states, shown using red and blue points. Red and blue curves denote parabolic fits to the photoemission data. These photoemission spectroscopy results are consistent with the first-principles calculations presented in panel **c**. A quantitative comparison between the two sets of data is provided in Extended Table I. The presence of band inversion and its related surface states indicate a topological insulator phase in α-As. Such a topological characterization relies on the presence of an energy gap between the occupied and empty bands at any given bulk momentum. However, it does not require the presence of an indirect gap between occupied and empty states across the entire Brillouin zone. Since, α-As satisfies the former but not the latter condition, it can realize a topological phase even though it is semi-metallic (its bulk bands cross the Fermi level).

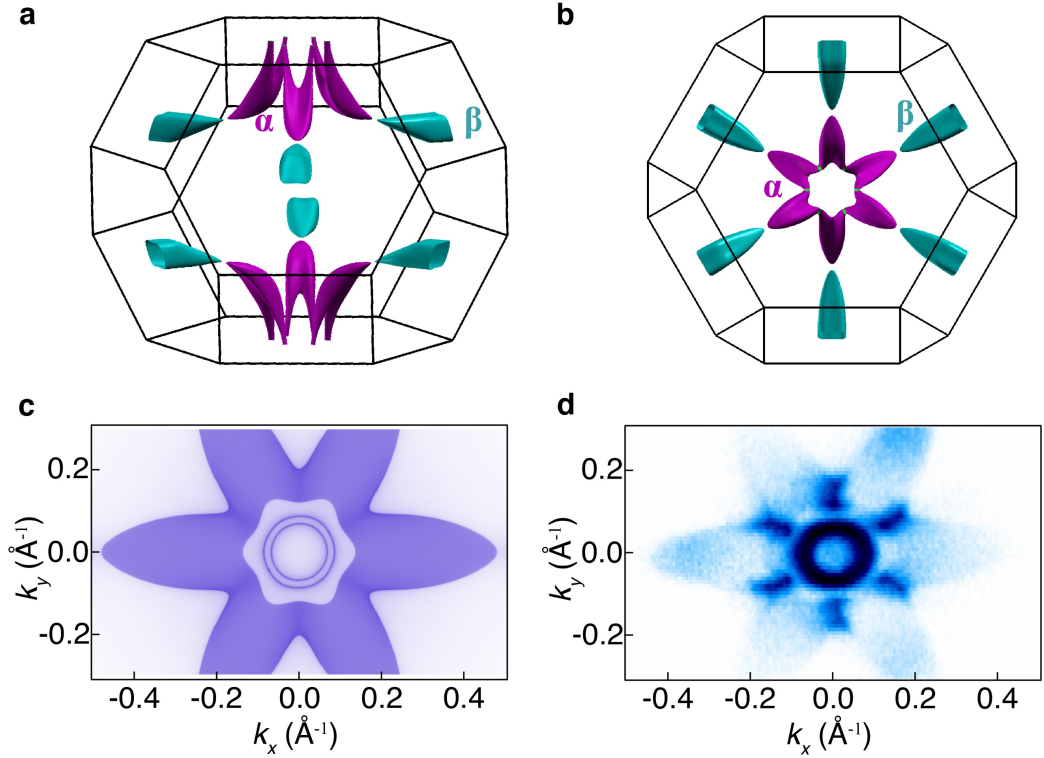

**Extended Fig. 2. Fermi surface of α-As obtained via first-principles calculations and photoemission spectroscopy. a and b**, Three-dimensional Fermi surface of α-As featuring a hole-like (α) pocket shown in magenta and an electron-like (β) pocket shown in cyan, originating from the magenta and cyan bands in Extended Fig. 1**b**, respectively. **c,** Fermi surface projected on the (111) surface obtained using first-principles tight-binding calculations. It exhibits two additional Fermi pockets centered at $k = 0$, due to the Rashba-like surface states. **e**, Fermi surface map taken on the As (111) surface. The map is obtained using 20 eV photon energy and with linear-horizontally polarized light. Photoemission data also show circular Fermi pockets centering at $k = 0$ in addition to the three-dimensional bulk state.

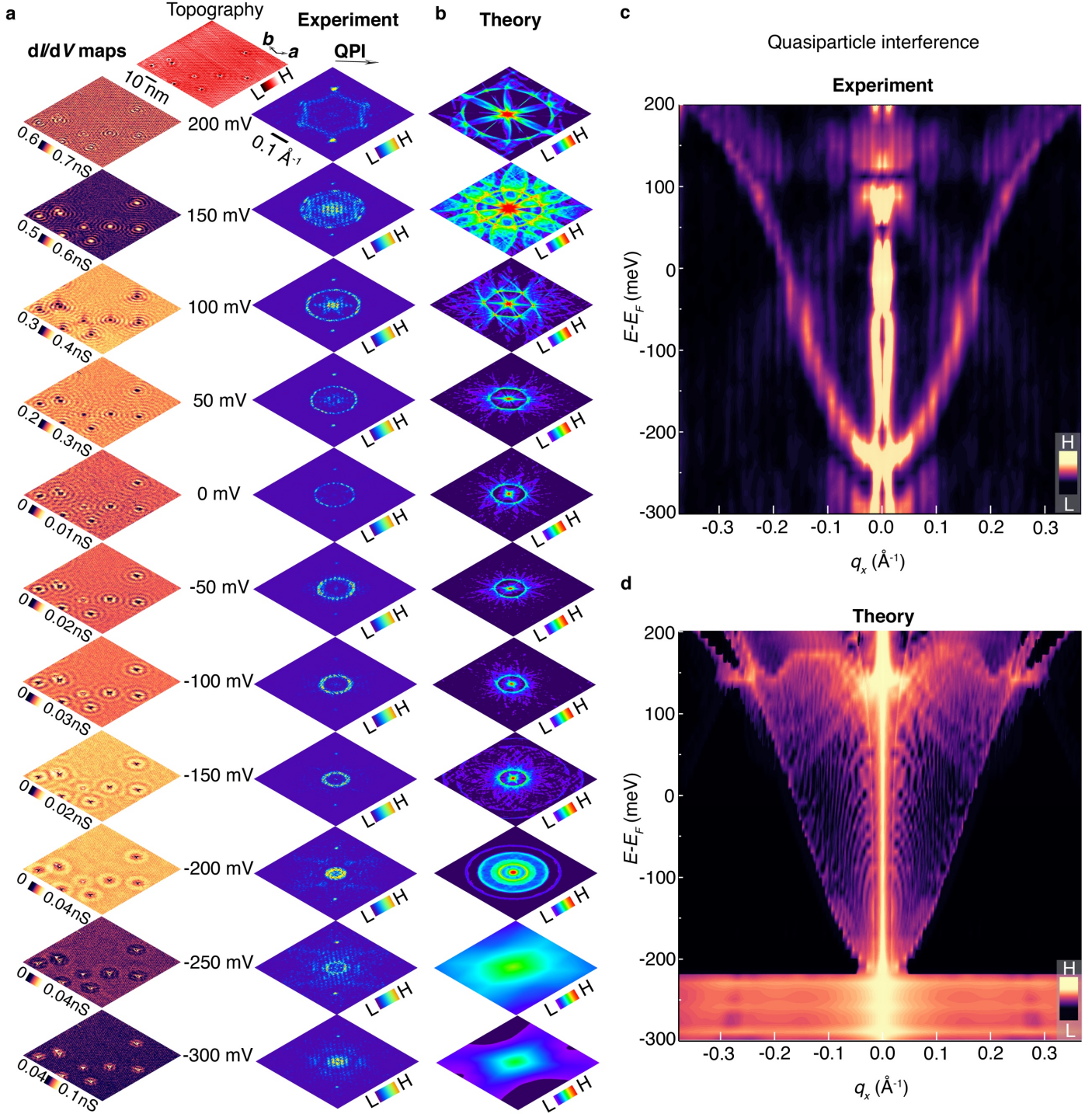

**Extended Fig. 3. Quasiparticle interference signature of the surface state. a,** Large-scale spectroscopic maps (left) and the corresponding Fourier transform at different tip-sample biases; the respective topography is also shown. Clear quasiparticle interference patterns are seen at all the bias voltages. **b,** Calculated quasiparticle interference patterns. The joint density of states method, which takes into account all possible scattering vectors, has been employed for the calculation. **c,** Experimental quasiparticle interference spectrum (energy resolution ≃ 15 meV which is determined by the number of bias points in the tunneling spectrum at each spatial point; $q$ resolution = 0.0015 Å$^{-1}$ which comes from the dimensions of the scanned area) taken along a high-symmetry direction, marked with an arrow in panel **a**, right column. The spectrum demonstrates several interference branches. The parabolic-shaped branch, which is the most prominent scattering branch in the quasiparticle interference spectrum, originates from $\sim -230 \pm 15$ meV. This energy position matches the location of the surface band bottom in the photoemission spectroscopy (Fig. 1**f**) and the first-principles calculation results (Fig. 1**g**). (A quantitative comparison between the three sets of data is provided in Extended Table I.) Therefore, it is most likely that this quasiparticle interference branch stems from the surface state. Tunneling junction set-up for d$I$/d$V$ maps: $V_{set}$ = 200 mV, $I_{set}$ = 1 nA, $V_{mod}$ = 5 mV. **d**, Calculated quasiparticle interference spectrum, showing a clear parabolic-shaped scattering branch, which is in excellent agreement with the experimental result presented in panel **c**.

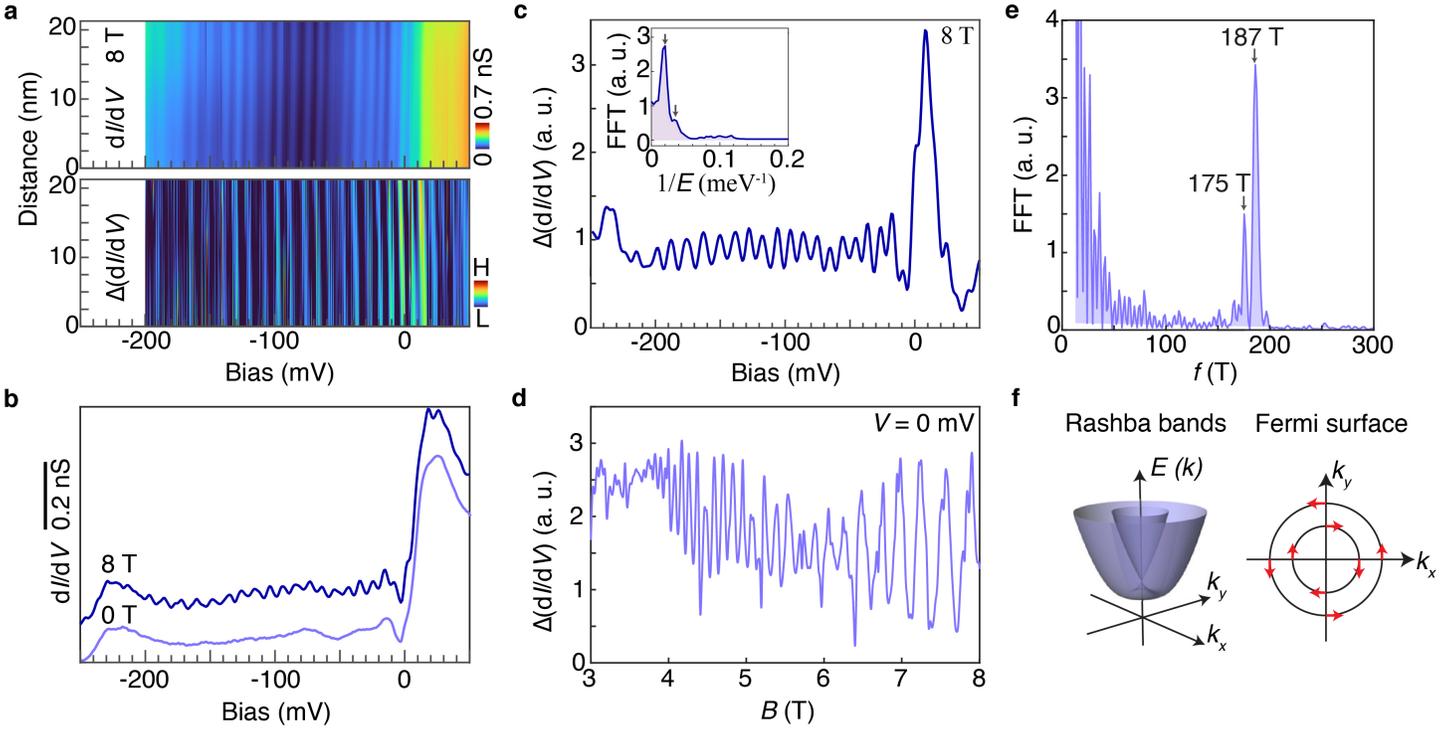

**Extended Fig. 4. Landau level spectroscopy of α-As. a**, d$I$/d$V$ line map (top) and its derivative (bottom) under $B$ = 8 T applied perpendicularly to the cleaved surface. The intense modulation in differential conductance is due to Landau quantization. **b,** Averaged d$I$/d$V$ spectra at $B$ = 0 T, and 8 T taken from the same region clarifying that the d$I$/d$V$ modulation is due to Landau quantization. **c,** Derivative (to remove the slowly varying background) of the 8 T data and its Fourier transform (inset) showing two (close-by) peaks, suggesting two sets of Landau fans. **d,** Field-dependent differential conductance data obtained from a line cut of the Landau fan (shown in Fig. 1**h**) at the Fermi energy. **e,** Fourier transform of panel **d** data (after replotting the data as a function of $\frac{1}{B}$). Fourier transform magnitude as a function of cyclotron frequency reveals two well-developed peaks at 175 T and 187 T. The two peaks correspond to the Fermi surface areas (in units of cyclotron frequency) of the two Fermi pockets stemming from the Rashba-split surface state. **f,** Schematic illustration of the Fermi surface stemming from the Rashba bands (see Extended Fig. 1**c** for numerically computed bands). Left: Schematic energy-momentum plot of the parabolic Rashba bands. A cut at the Fermi energy yields two concentric, circular Fermi pockets. Right: Schematic Fermi surface plot containing the two concentric circular pockets (see Extended Fig. 2**c** for numerically computed Fermi surfaces). The experimentally obtained 175 T and 187 T peaks (panel **e**) represent the area of the inner and outer circular Fermi pockets, respectively. Note that here the magnetic field is expected to open a gap at the surface state Dirac node which is located around −220 meV, far below the Fermi energy. Since there is no such crossing close to the Fermi energy, a magnetic field of 8 T is not expected to modify the Fermi surface significantly. Tunneling junction set-up: $V_{set}$ = 50 mV, $I_{set}$ = 0.5 nA, $V_{mod}$ = 0.5 mV.

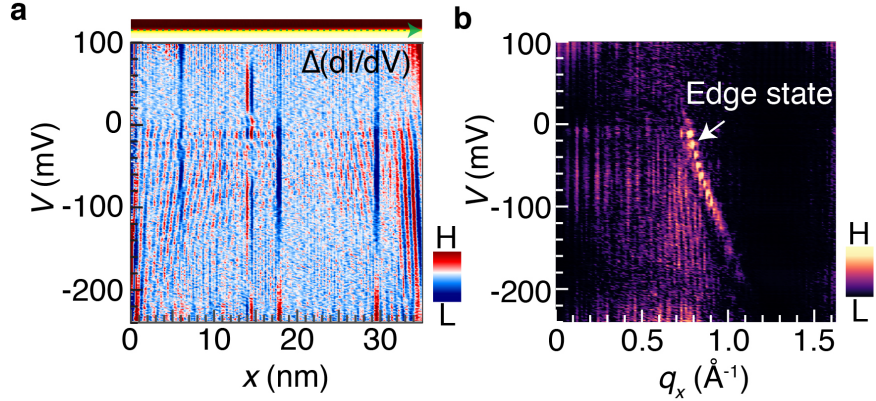

**Extended Fig. 5. Line spectroscopy with high spatial and energy resolutions and the corresponding Fourier transform taken along a bilayer step edge.** This bilayer step edge, whose orientation is identical to that of the step edges displayed in Figs. 3**a** and 3**d**, manifests a step edge state. **a**, Intensity plot revealing pronounced quantum interference patterns. The green dotted line on the topographic image (shown at the top) indicates the location where the d$I$/d$V$ line spectroscopy is performed. The direction of the scan is marked by an arrow. **b**, Corresponding one-dimensional Fourier transform, highlighting a clear edge state dispersion. Tunneling junction set-up: $V_{set}$ = 100 mV, $I_{set}$ = 0.5 nA, $V_{mod}$ = 0.5 mV.

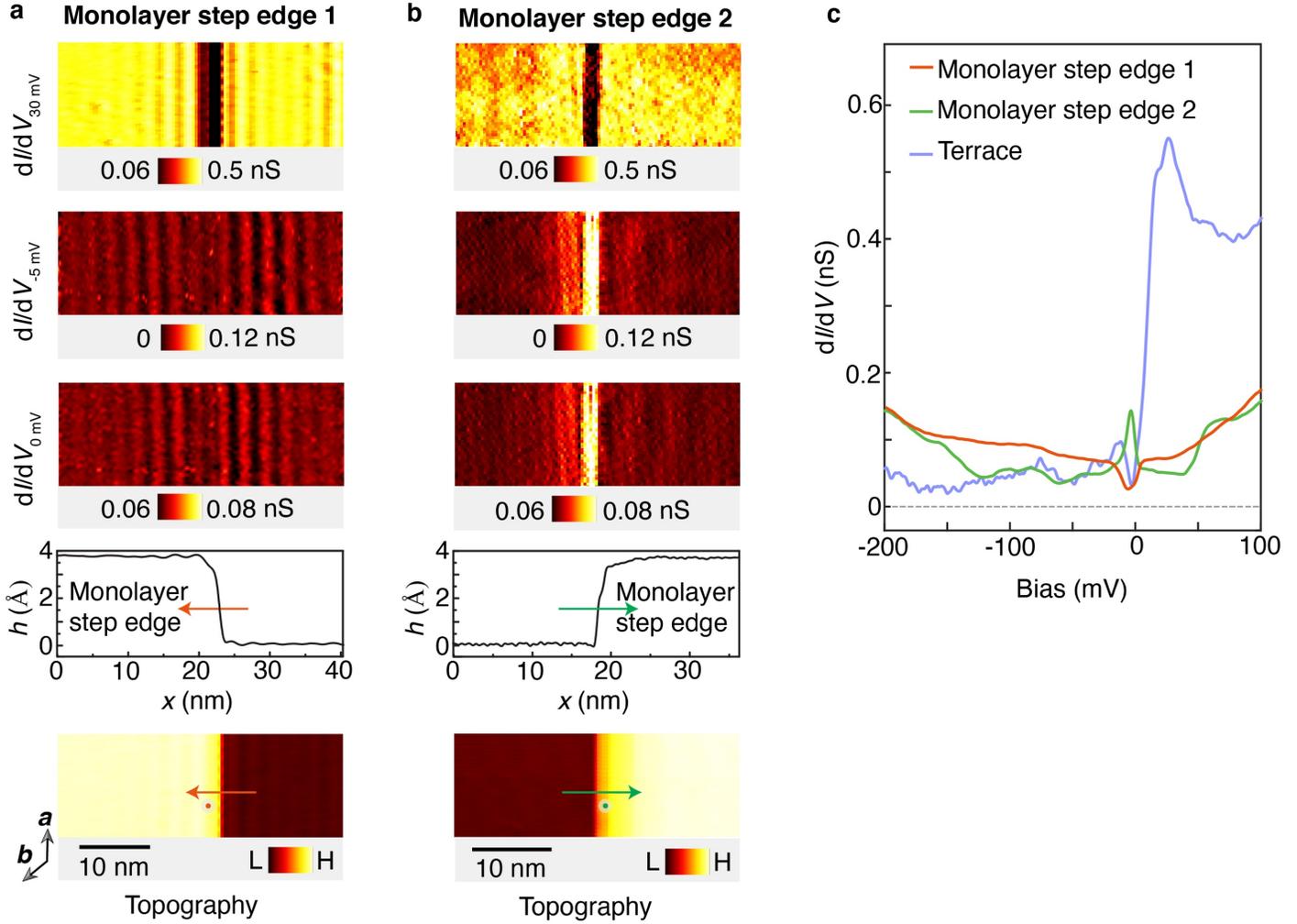

**Extended Fig. 6. Orientation-dependence of the monolayer step edge states. a, b,** Topographic images, height profiles, and the corresponding differential conductance maps around two monolayer step edges (1 and 2 as marked). The color-coded arrows in the topographies and the height profiles indicate the directions from the bottom to the top terraces. Depending on this direction, one monolayer step edge (monolayer step edge 2) exhibits a step edge state while the other monolayer step edge (monolayer step edge 1) does not. **c,** Differential spectra, taken at the monolayer step edge 1 (orange), monolayer step edge 2 (green), and away from the step edges (violet), revealing striking differences between the two step edges. Orange and green dots in the topographic images in panels **a** and **b** denote the respective positions on the monolayer step edges 1 and 2 where the differential spectra are taken. While the monolayer step edge 2 exhibits a pronounced step edge state, the monolayer step edge 1 features a largely suppressed differential conductance within the soft gap. Note that the monolayer step edge 2 has the same orientation as the monolayer step edge in Fig. 2**a**, and thereby both exhibit the same electronic structure. On the other hand, resembling the two bilayer step edges shown in Figs. 3**a, b**, the two monolayer step edges also show a dramatic contrast — one step edge exhibits a pronounced step edge state while the other step edge along the same crystallographic axis, but with a different orientation, does not. Here, however, the orientation favoring a step edge state is opposite to that of the bilayer case (Fig. 3**a, b**). Tunneling junction set-up for the differential spectra: $V_{set}$ = 100 mV, $I_{set}$ = 0.5 nA, $V_{mod}$ = 0.5 mV. Tunneling junction set-up for d$I$/d$V$ maps: $V_{set}$ = 100 mV, $I_{set}$ = 0.5 nA, $V_{mod}$ = 1 mV.

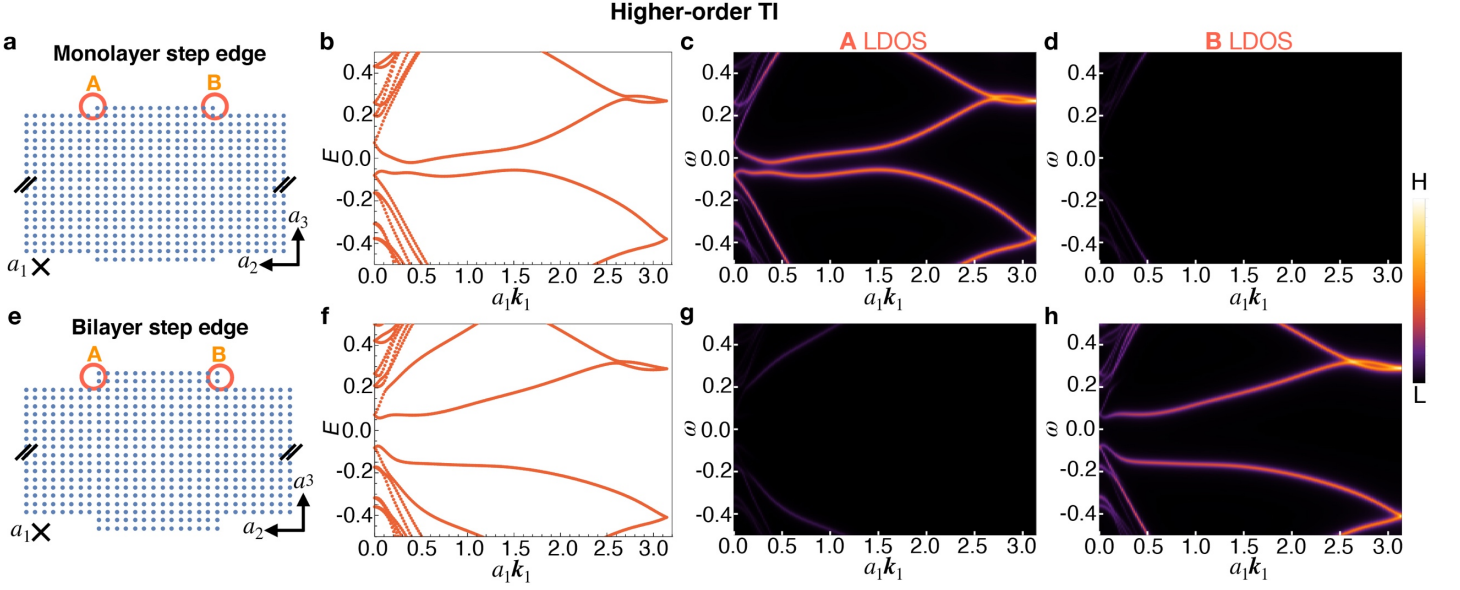

**Extended Fig. 7. Low-energy spectra of the tight binding model in absence of first-order topology.** Here we delineate the step edge dispersion in a tight-binding system that differs from α-As in that it only has a double band inversion at the $\Gamma$ point of the bulk Brillouin zone, and not an extra single band inversion at each of the three $L$, $L'$, $L''$ points as is the case for α-As. Correspondingly, the model only has nontrivial higher-order topology, but no first-order topology. As derived in our Supplementary Information, such a situation implies a gapped step edge dispersion, however, the gapped modes are precursors of hinge modes localized near the step edges. This situation is schematically depicted in Fig. 4c. **a,** Monolayer step edge geometry. We preserve periodic boundary conditions in the $a_1$-direction (the out-of-plane direction, with lattice spacing $a_1$), so that $k_1$ is a conserved crystal momentum. There are two step edges, one of type A and one of type B (see Supplementary Information), on each of the top and bottom surfaces. This is the minimal configuration of step edges that preserves inversion symmetry as well as periodic boundary conditions along the $a_2$-direction. **b,** Monolayer step edge dispersion with $k_1$ using the lattice shown in panel **a**. We only show the momentum range $k_1 \in [0, \pi]$ because the spectrum in the range $k_1 \in [\pi, 2\pi]$ is related by time-reversal symmetry. There is no nontrivial spectral flow (the spectrum is gapped). **c,** Local density of states (LDOS) for the A step edge. The LDOS is large for the low-energy bands closest to the gap, implying that they are well-localized at the step edge. **d,** Local density of states for the B step edge. This LDOS is small, implying that there are no low-energy states at the step edge. **e,** Bilayer step edge geometry. **f,** Bilayer step edge dispersion. There is still a gap, but the low-energy bands are slightly different from the monolayer case. **g,** Local density of states for the A step edge. The LDOS is again large for the low-energy bands, implying that they are well-localized at the same step edge as for the monolayer. **h,** Local density of states for the B step edge. Like the monolayer case, there are no step edge states close to the Fermi level.

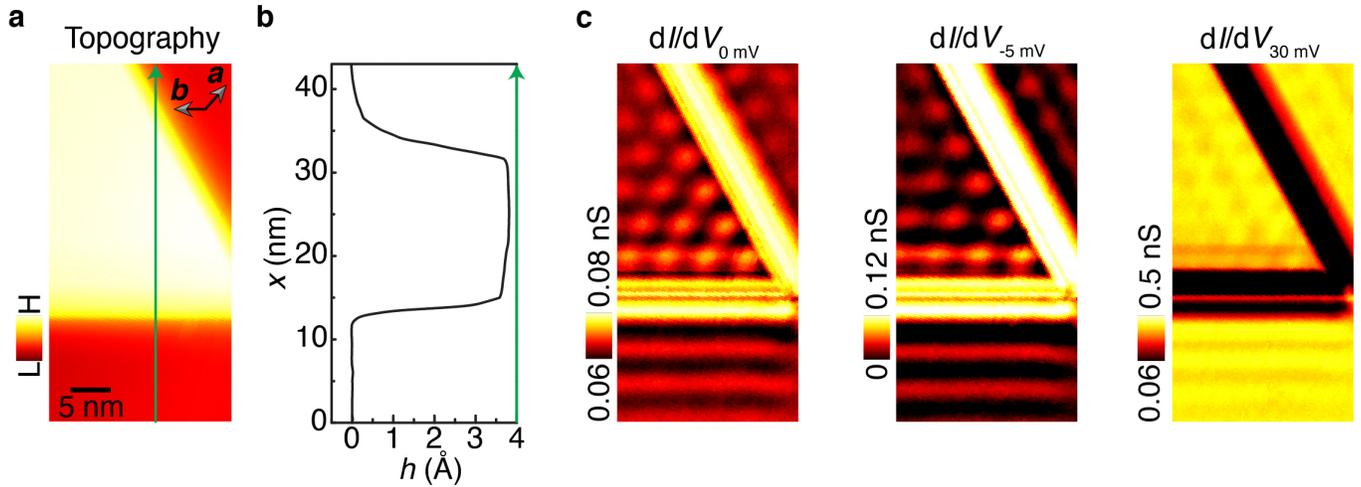

**Extended Fig. 8. Step edge states at the monolayer step edges along the *b*-axis and rotated by 120° with respect to it. a,** Topographic image of two monolayer step edges along the *b*-axis and rotated by 120° with respect to it. The orientation of the two step edges is opposite to those of the two monolayer step edges in Fig. 3**d**. Here the crystal side lies between the two step edges contrary to Fig. 3**d**, where the crystal side extends away from the pit. **b,** Height profile taken perpendicular to the *b*-axis direction. The corresponding location is marked on the topographic image in panel **a** with a green line; the direction of the scan is marked with an arrow. **c**, Differential conductance maps at different bias voltages, taken in the region shown in panel a. The two monolayer step edges, along the *b*-axis and rotated by 120° with respect to it, harbor step edge states that are manifested via an enhanced differential conductance. This behavior is in stark contrast to the monolayer step edges along the same directions, but with opposite orientations (Fig. 3**d**). Such an orientation dependence is consistent with our other data on different step edges along different crystallographic directions and with different orientations and layer thicknesses. Tunneling junction set-up for the differential spectra: $V_{set}$ = 100 mV, $I_{set}$ = 0.5 nA, $V_{mod}$ = 0.5 mV. Tunneling junction set-up for d$I$/d$V$ maps: $V_{set}$ = 100 mV, $I_{set}$ = 0.5 nA, $V_{mod}$ = 1 mV.

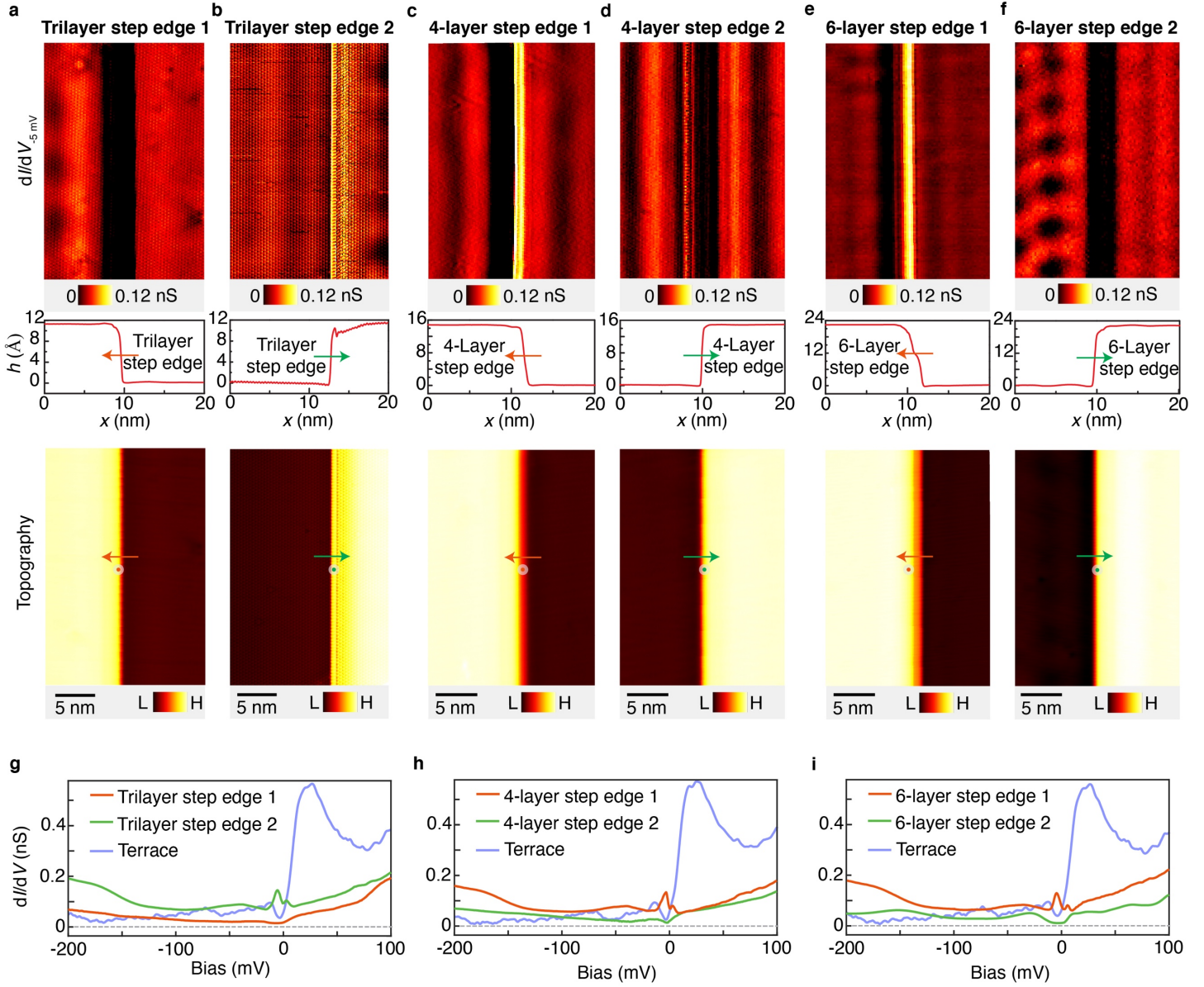

**Extended Fig. 9. Orientation-dependence of the three-, four-, and six-layer-thick step edge states.** Topographic images, height profiles, and the corresponding differential conductance maps at $V = -5$ mV around two tri-layer (**a** and **b**), four-layer (**c** and **d**), and six-layer (**e** and **f**) step edges. The color-coded arrows in the topographies and the height profiles indicate the directions from the bottom to the top terraces. Depending on this direction and the layer thickness, certain step edge orientations (such as tri-layer step edge 2, four-layer step edge 1, and six-layer step edge 1) exhibit step edge states, while other orientations (like tri-layer step edge 1, four-layer step edge 2, and six-layer step edge 2) do not. **g-i**, Differential spectra, taken at the two tri-layer (panel **g**), four-layer (panel **h**), and six-layer (panel **i**) step edges, revealing striking differences between the two step edge orientations. Consistent with the orientation dependence of the monolayer step edges in Extended Fig. 6, and opposite to that of the bilayer step edges in Figs. 3**a** and 3**b**, the tri-layer step edge 2 (green curve in panel **g**; d$I$/d$V$ map in panel **b**) shows a pronounced step edge state, while the tri-layer step edge 1 (orange curve in panel **g**; d$I$/d$V$ map in panel **a**) exhibits largely suppressed differential conductance within the soft gap of the surface. Interestingly, this orientation dependence switches for the four-layer step edge, where the four-layer step edge 2 (green curve in panel **h**; d$I$/d$V$ map in panel **d**) lacks a step edge state, while the four-layer step edge 1 (orange curve in panel **h**; d$I$/d$V$ map in panel **c**) carries a strong step edge state. This observation is consistent with the even-odd effect observed between the step-edge preferred orientations in mono and bilayer step edges. The six-layer step edge displays a similar geometry dependence as the four-layer step edge, with the six-layer step edge 2 (green curve in panel **i**; d$I$/d$V$ map in panel **f**) lacking a step edge state, while the six-layer step edge 1 (orange curve in panel **i**; d$I$/d$V$ map in panel **e**) exhibits a pronounced step edge state. Moreover, the data taken at the six-layer step edge, which is reasonably thick to allow for well-separated hinge states between the top and the bottom hinges, can be considered a potential candidate for the observation of the hinge state within the higher-order topological insulator phase. Tunneling junction set-up for the differential spectra: $V_{set}$ = 100 mV, $I_{set}$ = 0.5 nA, $V_{mod}$ = 0.5 mV. Tunneling junction set-up for d$I$/d$V$ maps: $V_{set}$ = 100 mV, $I_{set}$ = 0.5 nA, $V_{mod}$ = 1 mV.

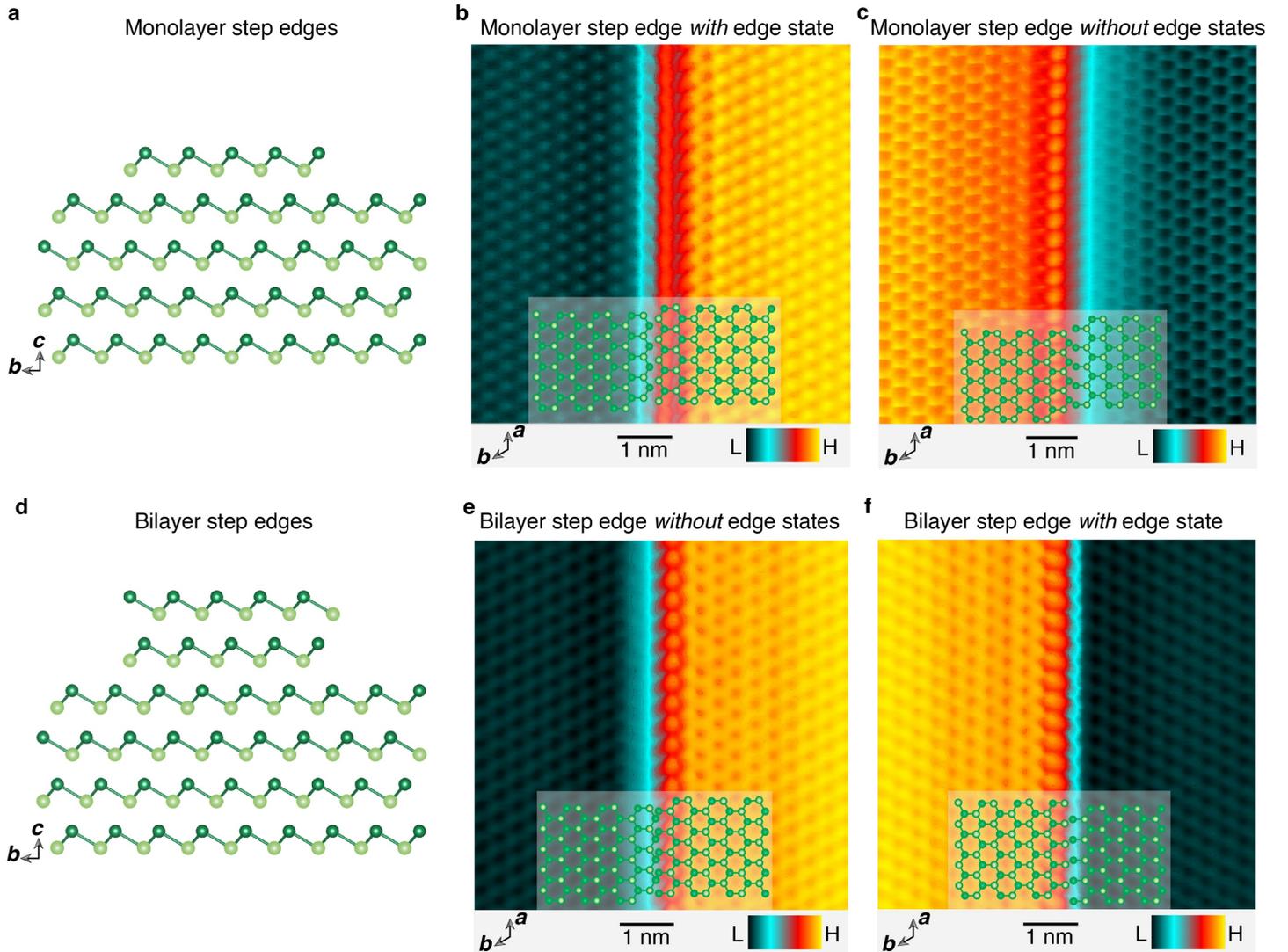

**Extended Fig. 10. Atomic structure for two different geometric orientations of the mono and bilayer step edges, highlighting the asymmetry between them. a**, Side view (along the *bc* plane) of the two monolayer step edges. **b** and **c**, Atomically resolved topographic images of the corresponding monolayer step edges along the *a*-direction. **d**, Side view (along the *bc* plane) of the two bilayer step edges. **e** and **f**, Atomically resolved topographic images ($V_{gap}$ = 100 mV, $I_t$ = 3

nA) of the corresponding bilayer step edges along the *a* direction. Notably, there is an asymmetry between the two orientations in both monolayer and bilayer step edges, with one orientation featuring a sharper step edge. Interestingly, this preferred orientation alternates between the mono and bilayer cases, in accordance with the even-odd effect observed in the step edge states, where the step edge state appears to favor the smoother edge. We note that visualizing atoms directly at the step edge poses challenges due to the abrupt change in height at that location and the small inter-atomic distance ($\simeq 3.7$ Å) in α-As. As a result, the atom arrays at the step edge may not be as clearly visualized in our scanning tunneling microscopy compared to those located away from the step edge.

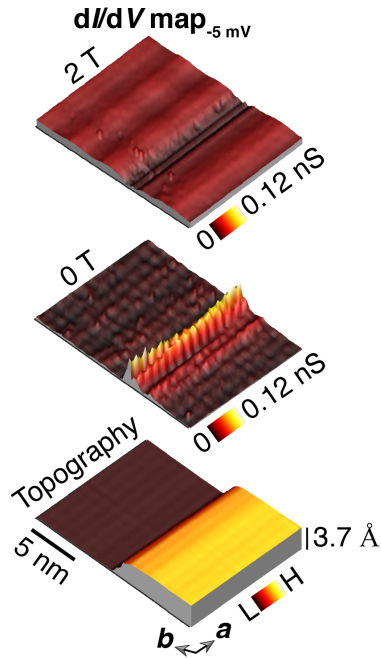

**Extended Fig. 11.** Topography (bottom) and the corresponding d$I$/d$V$ maps ($V = -5$ mV) acquired at $B = 0$ T and 2 T. At $B = 2$ T, the spectral weight at the step edge is suppressed. Such a suppression highlights the impact of time-reversal symmetry breaking on the step edge state. Tunneling junction set-up: $V_{set} = 100$ mV, $I_{set} = 0.5$ nA, $V_{mod} = 1$ mV.

**Extended Table I: Quantitative comparison between the surface state results obtained from photoemission spectroscopy, quasiparticle interference, and first-principles calculations. Error bars correspond to the experimental energy and momentum resolutions.**

| Surface state properties | Photoemission spectroscopy | Quasiparticle interference | First-principles calculations |
| --- | --- | --- | --- |
| Band bottom position | $-215 \pm 20$ meV | $-230 \pm 15$ meV | $-220$ meV |
| Rashba splitting, $\Delta k$ (For the state at the Fermi level) | $0.006 \pm 0.0025$ Å$^{-1}$ | N/A | $0.0056$ Å$^{-1}$ |
| Fermi wavevectors | 0.07± 0.0025 and 0.09± 0.0025 Å$^{-1}$ | $q_F = 2k_F \simeq 0.18\pm 0.0015$ Å$^{-1}$ | 0.075 and 0.093 Å$^{-1}$ |
| Fermi velocity | 2.46 eV·Å | $v_q = \frac{dE}{dq} = \frac{dE}{2dk} = \frac{v_k}{2} \simeq 1.6$ eV·Å | 2.48 eV·Å |

**Competing interests:** The authors declare no competing interests.

**Data and materials availability:** All data needed to evaluate the conclusions in the paper are present in the paper. Additional data are available from the corresponding authors upon reasonable request.